\pdfoutput=1

\documentclass[nofootinbib, showpacs, aps, prd, superscriptaddress, amsmath, amssymb, floatfix, a4paper, twoside,11pt]{revtex4-1}
\usepackage{dsfont}
\usepackage{graphicx}
\usepackage{bm}
\usepackage{color}
\usepackage{multirow}
\usepackage{bookmark}
\usepackage{xcolor}
\hypersetup{
    colorlinks,
    linkcolor={red!60!black},
    citecolor={blue!50!black},
    urlcolor={blue!80!black}
}

\newcommand{\vslash}[1]{#1 \hspace{-0.5 em} /}

\def	\g	{\gamma}
\def	\pd	{\partial}

\def	\e	{\varepsilon}
\def	\t	{\tau}
\def	\q	{\theta}
\def  \ll	{\parallel}
\def	\T	{\perp}
\def	\fr	{\dfrac}
\def  \~	{\approx}
\def	\(	{\left(}
\def	\)	{\right)}
\def	\x	{\cdot}

\def 	\eq	{
		
		\vspace{.2cm}
		\noindent}

\begin{document}
{\small
\begin{center}
\begin{tabular}{l r r}
Presented at IC Channeling 2018 &~~~~~~~~~~~&September 27, 2018 \\
Published on ArXiv &&  October 15, 2018\\
Accepted by JHEP  &&  March 12, 2019\\
Published on JHEP &&  March 26, 2019\\
\end{tabular}
\end{center}
}

\vspace{1cm}

\title{Feasibility of $\tau$-lepton  electromagnetic dipole moments measurement\\ using bent crystal at the LHC}

\author{A.S.~Fomin}
\email[]{alex.fomin@cern.ch}
\affiliation{NSC Kharkiv Institute of Physics and Technology, 1 Akademicheskaya St., Kharkiv, 61108 Ukraine}
\affiliation{currently at CERN, European Organization for Nuclear Research, Geneva 23, CH-1211 Switzerland}
\author{A.Yu.~Korchin}
\email[]{korchin@kipt.kharkov.ua}
\affiliation{NSC Kharkiv Institute of Physics and Technology, 1 Akademicheskaya St., Kharkiv, 61108 Ukraine}
\affiliation{V.N.~Karazin Kharkiv National University, 4 Svobody Sq., Kharkiv, 61022 Ukraine}
\author{A.~Stocchi}
\email[]{stocchi@lal.in2p3.fr}
\affiliation{LAL (Laboratoire de l'Acc\'el\'erateur Lin\'eaire), Universit\'e
Paris-Sud/IN2P3, Bâtiment 200 --- BP 34, Rue Andr\'e Ampere, 91898 Orsay
Cedex, France}
\author{S.~Barsuk}
\affiliation{LAL (Laboratoire de l'Acc\'el\'erateur Lin\'eaire), Universit\'e
Paris-Sud/IN2P3, Bâtiment 200 --- BP 34, Rue Andr\'e Ampere, 91898 Orsay
Cedex, France}
\author{P.~Robbe}
\affiliation{LAL (Laboratoire de l'Acc\'el\'erateur Lin\'eaire), Universit\'e
Paris-Sud/IN2P3, Bâtiment 200 --- BP 34, Rue Andr\'e Ampere, 91898 Orsay
Cedex, France}


\begin{abstract}

In this paper we discuss the possibility of measuring the anomalous magnetic and  
electric dipole moments of the $\tau$ lepton.
The method consists in studying the spin precession induced by the strong effective magnetic field inside 
channels of a bent crystal with a dedicated setup at the CERN Large Hadron Collider.\\

\noindent PACS number: 13.20.-v, 13.35.Dx, 13.40.Em, 13.88+e, 14.60.Fg, 61.85.+p\\

\noindent Keywords: Fixed target experiments, Polarisation, Tau Physics\\

\noindent ArXiv ePrint: \href{https://arxiv.org/abs/1810.06699}{1810.06699}\\

\noindent Cite this article as:

A.S.~Fomin, A.Yu.~Korchin, A.~Stocchi, et al. J. High Energ. Phys. 03 (2019) 156.

[arXiv:1810.06699]
\url{https://doi.org/10.1007/JHEP03(2019)156}

\end{abstract}

\maketitle

\tableofcontents

\vspace{1.cm}

\setcounter{footnote}{0}


\section{\label{sec:introduction} Introduction}

The magnetic dipole moment (MDM) of many particles has been measured \cite{PDG:2018}. 
For example, the experimental value of the electron MDM perfectly agrees with  theoretical calculations.   
For muon, there is a discrepancy between the Standard Model (SM) prediction \cite{PDG:2018} (ch.~57) 
and measurement by the BNL E821 collaboration~\cite{Bennett:2006fi}.
The deviation is about $3.5\,\sigma$ and it can possibly originate from effects of new physics (NP). 
This aspect of the muon MDM attracts considerable attention (see, {\it e.g.}, \cite{Knecht:2003, Passera:2004, Jegerlehner:2009}). 

The MDM of the $\tau$ lepton has not been measured so far and is of great interest 
for a test of the SM and the search for signatures of NP.

Usually one discusses the anomalous part of the MDM,
\eq
\begin{equation}
a_\tau = (g_\tau - 2)/2,
\label{eq:001}
\end{equation} 
which is expressed through the gyromagnetic factor $g_\tau$ (or $g$-factor). 
The value $g_\tau=2$ corresponds to a point-like Dirac particle ignoring radiative corrections. 
Deviations from $g_\tau =2$ arise from the known radiative corrections and can also stem from effects of NP. 
The latter are of the order of $m_\tau^2/\Lambda^2$, where $m_\tau$ is the mass of $\tau$ lepton and 
$\Lambda$ is the scale of NP. Because 
of the large mass of the $\tau$ lepton the effects of NP are expected to be more noticeable than the 
corresponding effects for muon.

The SM expectation for the $\tau$ MDM is given in Ref.~\cite{Eidelman:2007}, 
\eq
\begin{equation}    
a_{\tau, \, \rm{SM}} = 
	0.00117721(5).
         \label{eq:002}
\end{equation}

No measurement of the $\tau$ MDM has been performed because of the very short lifetime 
of the $\tau$ lepton, $2.903(5)\,\times\,10^{-13}\,{\rm s}$.  This does not permit to apply methods  
used in the electron and muon $g-2$ experiments. 
At present the $\tau$ anomalous magnetic moment $a_\tau$ is known to an accuracy of only about 
$10^{-2}$~\cite{PDG:2018}. This limit was obtained  by the DELPHI  
collaboration~\cite{Abdallah:2003} from the measurement of the $e^+ e^- \to e^+ e^- \tau^+ \tau^-$ 
total cross section at LEP2. The study of anomalous 
couplings of photon to $\tau$ lepton in this reaction at LEP was proposed in~\cite{Cornet:1995pw}.
At $95\,\%$ confidence level, the confidence interval is~\cite{PDG:2018}  
\eq
\begin{equation}
\label{eq:003}
    -0.052 < a_\tau  < 0.013.
\end{equation}
It is also given in Ref.~\cite{Abdallah:2003} in the form $a_\tau   =  -0.018(17)$. 
The authors of Ref.~\cite{GonzalezSprinberg:2000} analyzed LEP1, SLD and  LEP2 experiments 
on the $\tau$-lepton production and obtained $2 \sigma$ bounds
\eq
\begin{equation}
\label{eq:003a}
    -0.007 < a_\tau  < 0.005.
\end{equation}
Note that Ref.~\cite{GonzalezSprinberg:2000} assumed that the electric dipole moment (EDM) of the $\tau$ was absent, while Ref.~\cite{Abdallah:2003}, when obtaining bounds for $a_\tau$, assumed that EDM was equal to its SM value, which is negligibly small.  

In general, the EDM of the $\tau$ lepton, $d_\tau$,  can take nonzero values if both the parity $P$ and time reversal $T$ symmetries  
are violated \cite{Roberts:2010}. In contrast to the interaction of MDM with the magnetic field ${\cal \vec{B}}$ 
\begin{equation}
H_M = - \vec{\mu}_\tau \, {\cal \vec{B}} = - \frac{g_\tau}{2} \mu_B \, \vec{\sigma} {\cal \vec{B}},
\label{eq:003b}
\end{equation}
where $\mu_B= \tfrac{e \hbar}{2 m_\tau c}$ is the Bohr magneton,  
the interaction of EDM with the electric field ${\cal \vec{E}}$ 
\begin{equation}
H_E = -\vec{d}_\tau \,  {\cal \vec{E}} =  - \frac{f_\tau}{2}  \mu_B \, \vec{\sigma}  {\cal \vec{E}}, 
\label{eq:003c}
\end{equation}
is $P$, $T$ and also $CP$ violating.  This causes the interest to the EDM of leptons.  
In framework of the SM the lepton EDM are extremely small 
as being originated from the 4-loop diagrams and additionally because of the smallness of $CP$ violation via the CKM matrix 
\cite{Commins:1999, Commins:2009}. 
For example, the SM estimation for electron is $d_e \lesssim 10^{-38}$ e$\cdot$cm, 
while the $\tau$-lepton EDM can be larger by a factor $ \tfrac{m_\tau}{m_e}$, so that $d_\tau \lesssim 3.5 \cdot 10^{-35}$ e$\cdot$cm,   
or $f_\tau  \lesssim 1.2 \cdot 10^{-20}$.   
The currently achieved experimental accuracy~\cite{Inami:2002} cited by PDG \cite{PDG:2018},
\begin{eqnarray} 
\label{eq:003d}
&& -0.22 < {\rm Re} \, d_\tau < 0.45  \; \,  (10^{-16} \ {\rm e \cdot cm}),  \; \; {\rm or} \;   -0.008 < {\rm Re} \, f_\tau < 0.0162,
\\
&& -0.25 < {\rm Im} \, d_\tau \, < 0.008 \; \,  (10^{-16} \ {\rm e \cdot cm}),  \; \; {\rm or} \;   -0.009 < {\rm Im} \, f_\tau \, < 0.00029, 
\label{eq:003e}
\end{eqnarray}
is 18 orders of magnitude worse than the SM prediction. Clearly the SM value is hardly reachable 
in experiments and therefore any observation of the $\tau$ EDM will be indication of additional 
sources of $CP$ violation beyond the SM.

There have been various proposals to measure the MDM and EDM of the $\tau$ lepton. 
At the LHC, it was suggested in Refs.~\cite{Hayreter:2013, Hayreter:2015} to obtain constraints on the $\tau$ MDM 
in measurements of the Drell-Yan process.
The authors of Ref.~\cite{Atag:2010} discussed the possibility to set the limits on the $\tau$ MDM by 
studying the two-photon production of $\tau$ leptons, $\gamma \gamma \to \tau^+ \tau^-$.  
The process $p \,p \to p \, \gamma^\star \, p \to p \, \tau^- \, \bar{\nu}_\tau \, 
q^\prime \, X$  with almost real photon $\gamma^\star$ is examined in Ref.~\cite{Koksal:2017nmy}. 

In Refs.~\cite{Koksal:2018env,Ozguven:2016rst} the potential of future 
colliders for improvement of the current experimental bounds is investigated. In particular,
at the electron-positron linear collider CLIC in the processes 
$\gamma \, \gamma \to \tau^+ \,  \tau^-$, \ $\gamma \, \gamma \to \tau^+ \, \tau^- \, \gamma$,   
and in the reactions $e^- \, e^+ \to e^- \, \gamma^\star \, e^+ \to \nu_e \, \tau^-  \, \bar{\nu}_\tau \, e^+$ 
with quasi-real photon 
and $ e^- \, \gamma \to \nu_e \, \tau^- \, \bar{\nu}_\tau$ with Compton back-scattering photon. 
The reaction $ e^- \,  p \to  e^- \, \gamma^\star \, \gamma^\star \, p \to e^- \tau^- \, \tau^+ \, p$  
on the future electron-proton colliders LHeC and FCC-he with high center-of-mass energies and high 
luminosities is considered in \cite{Koksal:2018xyi}.    

The authors of Refs.~\cite{Bernabeu:2008, Bernabeu:2009} proposed to determine the $\tau$ Pauli form factor 
$F_2 (s)$  from the electron--positron annihilation to a pair of tau leptons at the high-luminosity $B$ factory 
Super-KEKB. The process \mbox{$e^+ e^- \to \tau^+ \tau^-$} is planned to be studied at the $\Upsilon$ resonances  
 $\Upsilon(1S), \, \Upsilon(2S), \, \Upsilon(3S)$.  
This idea was further reconsidered and refined in Ref.~\cite{Eidelman:2016}.       

A new  method was proposed in Refs.~\cite{Eidelman:2016,Fael:2016} to probe the $\tau$ dipole moments from 
radiative leptonic $\tau$ decays  $\tau^- \to \ell^- \, \nu_\tau \, \bar{\nu}_{\ell} \, \gamma$  ($\ell = e, \, \mu$) at the  
KEKB and Super-KEKB colliders.
The feasibility studies~\cite{Eidelman:2016} show that such a measurement at Belle II will be competitive with the
 current limits for MDM given in Eq.~(\ref{eq:003}). 

Another method to determine the MDM of the $\tau$ lepton was discussed in Ref.~\cite{Samuel:1991}. It is based 
on the phenomenon of the particle spin precession in a bent crystal. This method has been successfully used by 
the E761 collaboration at Fermilab~\cite{Chen:1992} to measure the MDM of the $\Sigma^+$ hyperon. 
The authors of \cite{Samuel:1991} discussed the decay $B^+ \to \tau^+ \nu_\tau$ as a source of polarised $\tau$ 
leptons. However, the branching fraction of this decay is only about $10^{-4}$ which can make realization of this 
idea practically difficult.  

Nevertheless, the method which uses channeling of short-lived baryons in bent crystals has recently attracted
 considerable attention~\cite{Baryshevsky:2016,Fomin:2017,Botella:2016, Bagli:2017,Baryshevsky:2017}.
In particular, in Ref.~\cite{Fomin:2017} the feasibility of measuring the MDM of the charm baryons 
$\Lambda_c^+$ and $\Xi_c^+ $ at the LHC was addressed (see also details in Ref.~\cite{FominThesis}).
The authors of \cite{Botella:2016, Bagli:2017} studied the application of this method for future measurements of 
EDM and MDM of the strange, charm and beauty baryons.

In the present paper we would like to propose the idea of measuring the MDM and possibly EDM of the $\tau$ lepton through its spin precession in a bent crystal.
We suggest to use polarised $\tau$ leptons which come from the decay of the $D_s$ mesons produced at the LHC.
Because of the quark structure of $D_s$ meson ($D_s^+ = c \bar{s}$), the branching fraction of interest is quite sizeable,  ${\cal B}(D_s^+ \to \tau^+ \nu_\tau)  = 0.0548(23)$~\cite{PDG:2018}.  

The $D_s$ mesons are produced in $pp$ collisions with very high energies, of a few TeV, and subsequently 
decay to polarised $\tau$ leptons.
The leptons can be directed into a bent crystal, get in the channeling regime, and the direction of 
$\tau$ polarisation after the precession in the crystal can be determined from the angular analysis of its decay 
products. Schematically, the whole process is 
\eq
\begin{equation}
p\,p \to D_s^+ X, \quad
D_s^+ \to \tau^+ \nu_\tau, \quad
\tau^+ \ {\rm flight\ in\ the\ crystal}, \quad
\tau^+ \to  \pi^+\pi^+\pi^-\bar{\nu}_\tau.  
\label{eq:00}
\end{equation}

For the present study we consider the $\tau$ decay into three charged pions, which allows to identify the $\tau$ 
lepton through the reconstruction of a secondary vertex. The various  aspects of this measurement are studied in 
the present paper.

\newpage



\section{\label{sec:principle} Principle of the measurement}

\subsection{\label{subsec:Initial Polarisation} Initial polarisation of $\tau$ leptons from $D_s$ decays}

In $pp$ collisions the $\tau$ leptons are mainly produced in the decay   
$D_s^+  \to \tau^+ \nu_\tau$,
with a polarisation vector (see details in Appendix~\ref{app:1})
\eq%
\begin{equation}
\vec{\mathds P} = \frac{1}{M_D {k}_\tau^* } \left[ m_\tau \vec{p}_D \, - \, E_D \vec{k}_\tau \, + \, 
\frac{\vec{k}_\tau \left(\vec{k}_\tau \cdot \vec{p}_D\right)}{m_\tau+\varepsilon_\tau}\right]. 
\label{eq:006}
\end{equation}
Here $(E_D, \, \vec{p}_D)$ is the four-momentum of $D_s$ meson, $(\varepsilon_\tau, \, \vec{k}_\tau)$ 
is the four-momentum of $\tau$ lepton in the lab frame, and ${k}_\tau^* = (M_D^2 - m_\tau^2)/(2M_D)$ is 
the momentum of $\tau$ lepton in the rest frame of the $D_s$ meson.

It is convenient to separate the longitudinal and transverse components of the polarisation vector with respect 
to the direction of motion of the $\tau$ lepton, 
\eq%
\begin{eqnarray}
&& \vec{\mathds P}_{\parallel} = \frac{1}{M_D {k}_\tau^*} \, ( \varepsilon_\tau   p_D \cos \theta - E_D k_\tau) \, \vec{n}_k, \nonumber \\ 
&& \vec{\mathds P}_{\perp} = \frac{m_\tau p_D}{M_D {k}_\tau^*} \, ( \vec{n}_p - \vec{n}_k \cos \theta ),
\label{eq:007}
\end{eqnarray} 
where we defined unit vectors $\vec{n}_k \equiv \vec{k}_\tau/ {k}_\tau$ and $\vec{n}_p \equiv \vec{p}_D / {p}_D$, 
and $\cos \theta = \vec{n}_k \vec{n}_p$.   
These components are not independent and satisfy  
\eq%
\begin{eqnarray}
\label{eq:008}
&& \vec{\mathds P}_{\parallel}^2 + \vec{\mathds P}_{\perp}^2 =1, \\ 
&&  |\vec{\mathds P}_{\perp}| =  \frac{m_\tau {p}_D}{M_D {k}_\tau^* } \sin \theta.  
\label{eq:009}
\end{eqnarray}

\begin{figure}[t]
\begin{center}
\includegraphics[width=0.69\textwidth]{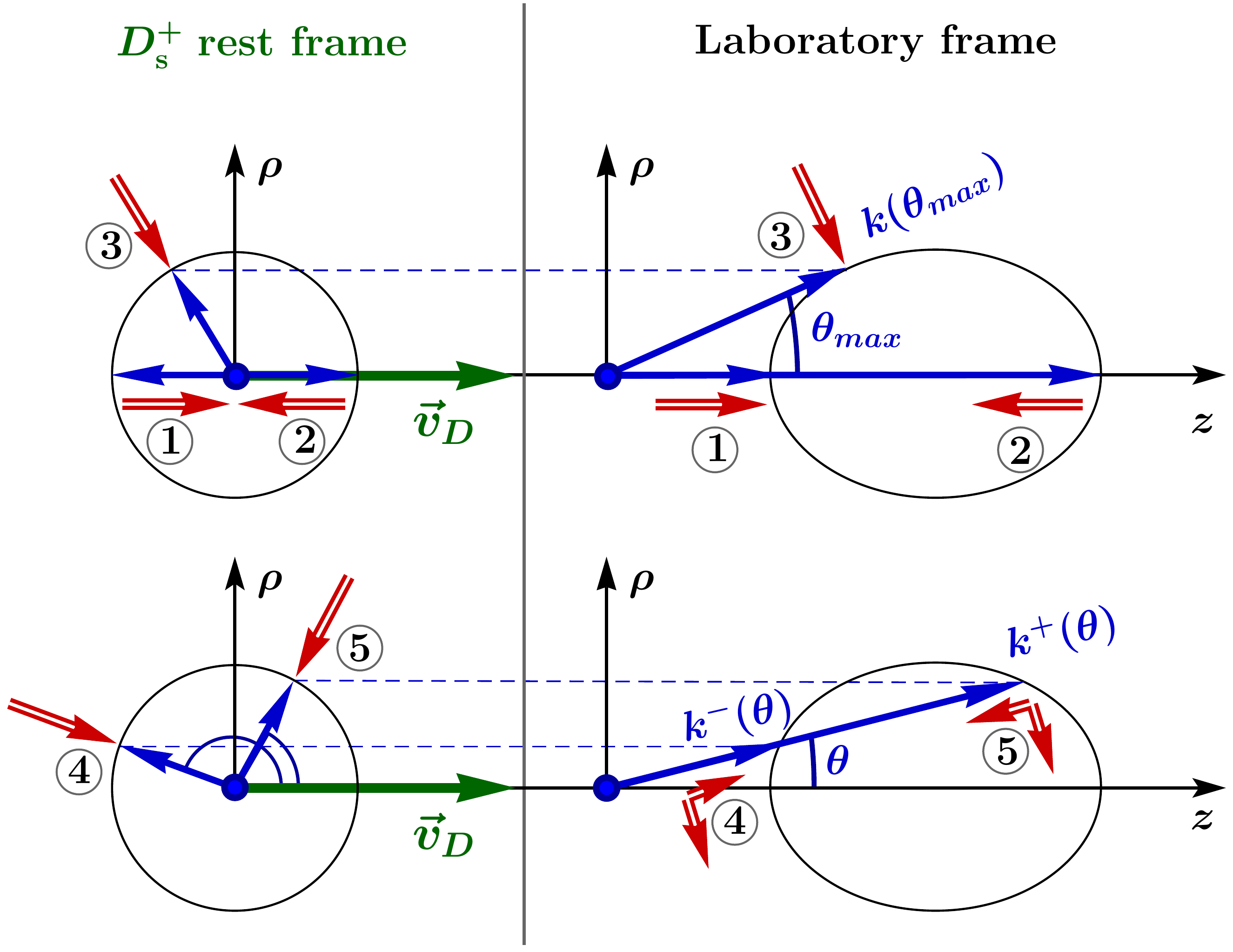}
\end{center}
\caption{The decay $D_s^+ \to \tau^+ \nu_\tau$ in the (left) $D_s$ rest frame and (right) Lab frame.
The axis $Oz$ is chosen along the direction of $D_s$ momentum.
Rotational symmetry around the $Oz$ axis is implied.
The ellipsoid is actually very elongated along the axis $Oz$ because  $\gamma_D \gg 1 $.
The blue arrows show the $\tau$ momentum $\vec{k}_\tau$, and the thick red arrows the $\tau$ polarisation 
vector,
$\vec{\mathds P}$.
Several cases of  $\tau$ polarisation in the Lab frame are shown:
(1,2) $\q=0$, (3) $\q=\q_{max}$ and (4,5) an arbitrary $\q$. 
}
\label{fig:frame}
\end{figure}

It is seen from Eqs.~(\ref{eq:008}) and (\ref{eq:009}) that if the $\tau$ lepton moves in the direction of 
the $D_s$ meson, i.e. $\theta =0$, then the polarisation is purely longitudinal and 
$|\vec{\mathds P}_{\parallel}| = 1$. The transverse polarisation increases with increasing $\theta$ and 
takes its maximal value $|\vec{\mathds P}_{\perp}|=1$ at the maximal allowed angle determined from
\eq%
\begin{equation}
\sin \theta_{max} = \frac{M_D {k}_\tau^* } {m_\tau {p}_D} = \frac{\gamma_\tau^* v_\tau^*}{\gamma_D v_D},
\label{eq:010}
\end{equation}
where $\vec{v}_D=\vec{p}_D/E_D$ is velocity of the $D_s$ meson in the lab frame, 
$\vec{v}_\tau^*=\vec{k}_\tau^*/\varepsilon_\tau^*$ is velocity of the $\tau$ lepton in the rest frame of 
the $D_s$ meson, and the corresponding Lorentz factors are $\gamma_D=E_D/M_D$ and 
$\gamma_\tau^* = \varepsilon_\tau^*/m_\tau$. 
Figure~\ref{fig:frame} shows schematically the ellipsoid of the $\tau$ momenta and the 
behavior of the $\tau$ polarisation.

At the minimal $\tau$ energy, 
\eq%
\begin{equation}
\varepsilon_\tau^{-} (\theta= 0) = \frac{1}{M_D} (E_D \varepsilon_\tau^* - p_D k_\tau^*)
\approx \frac{m_\tau^2}{M_D^2} E_D,  
\label{eq:e-_t}
\end{equation}
the polarisation is longitudinal and $\vec{\mathds P}_{\parallel} = + \vec{n}_z$, while at the maximal $\tau$ energy,   
\eq%
\begin{equation}
\varepsilon_\tau^{+} (\theta= 0) =  \frac{1}{M_D} (E_D \varepsilon_\tau^* + p_D k_\tau^*)
\approx E_D,
\label{eq:e+_t}
\end{equation}
the polarisation is longitudinal and $\vec{\mathds P}_{\parallel} = - \vec{n}_z$.
At the $\tau$ energy corresponding to the maximal angle $\theta_{max}$,
\begin{equation}
\varepsilon_\tau^{+} (\theta_{max})= \varepsilon_\tau^{-} (\theta_{max})=
\frac{2 m_\tau^2 }{M_D^2 + m_\tau^2} E_D, 
\label{eq:012}
\end{equation}
the polarisation is purely transverse and $|\vec{\mathds P}_{\perp}(\theta_{max})| =1$.

We can rewrite the expressions for the longitudinal and transverse $\tau$ polarisations, Eq. ~\eqref{eq:007},
as functions of the $\tau$ and $D_s$ energies, assuming $\e_\t=k_\t$, as we consider only energetic $\tau$
leptons,
\eq%
\begin{eqnarray}
&&\vec{\mathds P}_\ll( E_D, \e_\t) 
	    \~	\   -  \,  	\frac{	\,\,\cfrac{M_D^2+m_\t^2}{2\,m_\t^2}  \,  -  \,  \cfrac{ E_D}{\e_\t}
				\,\,}{   \cfrac{M_D^2+m_\t^2}{2m_\t^2}   \,   -  \,  1  }
		\,\vec{n}_k,  \nonumber \\
&&|\vec{\mathds P}_\T( E_D, \e_\t)|	\~	\fr{ \,2 \, m_\t^2  }{  M_D^2 \, - \, m_\t^2  }
					\sqrt{	 \(	\fr{M_D^2}{m_\t^2}
						     -	\fr{ E_D}{\e_\t}
						 \) \(	\fr{ E_D}{\e_\t} \, - \, 1 
					\) }.
\label{eq:Ptr( E_D)}
\end{eqnarray}
\eq%

\begin{figure}[b]
\begin{center}
\includegraphics[width=0.49\textwidth]{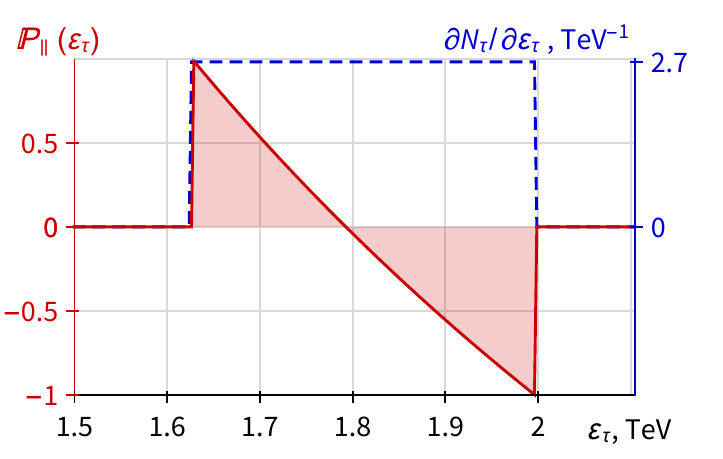}
\includegraphics[width=0.49\textwidth]{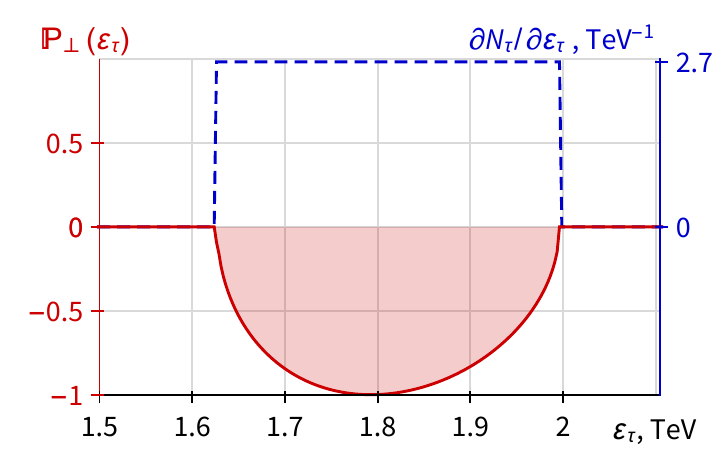}
\end{center}
\caption{
(Left) Longitudinal and (right) transverse polarisations as a function of the energy of $\tau$ leptons produced by 
$2\,$TeV $D_s$ mesons. The red solid line is the polarisation and the blue dashed line is the $\tau$ spectrum.
}
\label{fig:P_2TeV}
\end{figure}

Figure~\ref{fig:P_2TeV} presents the transverse and longitudinal polarisations of Eq.~\eqref{eq:Ptr( E_D)} of 
$\tau$ leptons produced by $D_s$ mesons of energy $E_D=2\,$TeV, as a function of  the $\tau$ energy.

The blue dashed line shows the $\tau$ spectra, which in the relativistic case is a step function.
Due to a small asymmetry (see Fig.~\ref{fig:P_2TeV}, left) there is a small average longitudinal polarisation of the order of $10\,\%$ in the direction $-\vec{n}_z$.

Note, that due to a cylindrical symmetry, $\tau$ leptons produced from $D_s$ decays of the same energy 
at the same polar angle $\q$ but with opposite azimuthal angles,  $\phi$ and $\phi+\pi$, 
would have an almost opposite transverse polarisation,
\eq%
\begin{equation}
\vec{\mathds P}_{\perp}(\theta,\phi) \approx -\vec{\mathds P}_{\perp}(\theta,\phi+\pi).
\label{eq:opposite}
\end{equation}
Thus, for the sufficient precision at the $\tau$ polarisation analysis, the directions of the $D_s$ and $\t$ momenta should be known with a precision better than $\q_{max}$.


\subsection{\label{subsec:Final Polarisation} Final polarisation and energy of reconstructed $\tau$ leptons}

Let us discuss possibilities to measure the $\tau$ polarisation. In general, it can be obtained from the angular distribution of the final particles in the $\tau$ decays. 
In a two-body decay, this distribution has the form
\eq%
\begin{equation}
\frac{1}{N} \frac{d N}{d \Omega} = \frac{1}{4\pi} \left( 1 + \alpha\,\vec{\mathds P}\cdot \vec{n} \right),
\end{equation}
 where $\vec{n}\equiv \vec{p}/{p}$ is the unit vector in the direction of movement of the final particle-analyzer and $\alpha$ is the asymmetry parameter.   
Several hadronic decay channels can be considered. They are listed in Table~\ref{tab:tau_decays}.

\begin{table}[!t]
\caption{Several $\tau$ hadronic decay channels  with asymmetry parameter, $\alpha$, and sensitivities. 
Sensitivities, $S$, are shown for the measurements made with different number of observables, 
$\vec{n}_k$ is the direction of $\tau$ momentum.  Columns 2,3 and 4 are our calculations and 
columns 5, 6, 7 and 8  are from the analysis of Ref.~\cite{Davier:1993}.  }
\begin{center}
\begin{tabular}{ |l |c |c | c | c | c | c | c|}
  \hline
~$\tau$ decay  &  $\alpha$ & $\alpha_{\rm eff}$ &
$S(1)_{\rm eff}$ & S(1)  &  $S(2)$  & ~S(all but $\vec{n}_k$)~ & ~S(with $\vec{n}_k$)~ \\
	\hline
~$\tau^+ \to \pi^+ \bar{\nu}_\tau$ &
 1  & 1 & 0.58  & 0.58 &  -- & 0.58 & 0.58  \\  
~$\tau^+ \to \rho^+ \, (2 \pi) \, \bar{\nu}_\tau $ &
0.45  & 0.49 & 0.28  & 0.26  & 0.49 & 0.49 & 0.58 \\  
~$\tau^+ \to a_1^+ \, (3\pi) \, \bar{\nu}_\tau~$ &
~0.021~ & ~0.12\,--\,0.21~ & ~0.069\,--\,0.12~  & ~0.10~ & ~0.23~  & 0.45 & 0.58 \\  
\hline
\end{tabular}
\end{center}
\label{tab:tau_decays}
\end{table}

In the second column, the parameter $\alpha$ for the two-body decay is given. 
Other columns contain information on the sensitivity 
$S = 1/(\sigma \sqrt{N})$ to the polarisation, where $N$ is the number of events, $\sigma$ is the error on 
polarization measurement at $\vec{\mathds P}=0$. According to Ref.~{\cite{Davier:1993}} 
if the full kinematics of the decay is reconstructed then the maximum possible sensitivity $S=1/\sqrt{3} \approx 0.58$ is  reached. 

For the single-pion channel, the maximum sensitivity is achieved, even though this channel suffers 
from large background. 
For the vector-meson channels, the sensitivity is suppressed because of the smaller asymmetry parameter 
$\alpha = ({m_\tau^2 - 2 m_v^2})/({m_\tau^2 + 2 m_v^2})$  \ ($m_v =  m_\rho, \, m_{a_1}$).  
The parameter $\alpha$ for the $a_1 (1260)$ meson turns out to be 0.021 which leads to the low sensitivity about  0.01. 
However we can estimate the influence of the large decay width of this resonance.  For this, we integrate $\alpha$ 
as a function of the invariant mass 
$W$ of a resonance, weighted with a normalised Breit-Wigner distribution,
\eq%
\begin{equation}
\alpha_{\rm eff} =  \int_0^{m_\tau} \left( \frac{m_\tau^2 - 2 W^2}{m_\tau^2 + 2 W^2} \right) \,
\frac{{\rm d}W}{( W^2 - m_{v}^2)^2 + m_{v}^2 \Gamma_{v}^2} \, \left( 
\int_0^{m_\tau} \frac{{\rm d}W}{( W^2 - m_{v}^2)^2 + m_{v}^2 \Gamma_{v}^2} \right)^{-1}.
\end{equation}  
Choosing $m_{a1}=1230$ MeV and the decay width $\Gamma_{a1} = 250 - 600 $ MeV \cite{PDG:2018},
we obtain an interval for the effective parameter for the $a_1 (1260) \, \nu_\tau$ mode in the third column of
Table~\ref{tab:tau_decays}. The corresponding sensitivity is given in the fourth column.  
It is seen that accounting for the large decay width of $a_1 (1260)$ enhances the sensitivity considerably. 
For the $\rho \, \nu_\tau$ mode, the effect of the width is much less important as seen also in 
Table~\ref{tab:tau_decays}.

The analysis of Ref.~\cite{Davier:1993} shows that the sensitivity of the polarisation measurement increases 
with increasing the number of observables, and reaches the maximal value of 0.58 if, in addition, 
the direction of the $\tau$ momentum is determined. 
Thus one can opt for the $3-$prong decay $\tau^+ \to a_1^+ \, \bar{\nu}_\tau  \to \pi^+ \pi^+ \pi^- \, \bar{\nu}_\tau $ 
as it can be separated from background processes and allows us to reconstruct the position of the vertex. 
Note that various aspects of the $\tau \to 3 \pi \nu_\tau$ decay mode are studied in Refs.~\cite{Hagiwara:1989fn, Rouge:1990, Duflot:1993}.  

In this paper we do not address the reconstruction of $\tau$ polarisation, but an estimation on sensitivity.
The sensitivity on polarisation in $\tau^+\to a_1^+ \overline\nu_\tau$ decay is very small $S \sim 0.1$. On the other hand, if we also analyse the consequent decay $a_1^+\to \pi^+\pi^+\pi^-$ the sensitivity reaches its maximal possible value $S \sim 0.58$. Thus, a big part of sensitivity comes from $a_1$, which can be fully reconstructed, so we can safely estimate that $S>0.2$.

For this reason in the sensitivity studies (in Sec.~\ref{sec:sensitivity}) we chose to use the range for sensitivity on polarisation $(0.2<S<0.58)$.

The energy of $\tau$ can be partially reconstructed. Assuming that we fully reconstruct $a_1$ from $a_1^+\to \pi^+\pi^+\pi^-$ decay, using the relation between energies of $a_1$ and $\tau$, that is analogous to the ones shown in Eq.~\eqref{eq:e-_t} and \eqref{eq:e+_t}, we obtain the relative error on $\tau$ energy
\eq%
\begin{equation}
\frac{\Delta\varepsilon_\tau}{\varepsilon_\tau}=
\frac{1}{\,\sqrt{3}\,}\ 
\frac{\ m_\tau^2/m_{a_1}^2-1\,}{m_\tau^2/m_{a_1}^2+1}\approx0.21.
\label{eq:dE}
\end{equation}

Of course, putting the additional silicon telescope detector in front of the crystal could allow to fully reconstruct the decay kinematics, but this setup would require a detailed study of the background in the environment where the experiment will be located.

\newpage


\subsection{\label{subsec:Ds} Experimental setup}

When a relativistic $\tau$~lepton is deflected by a bent crystal, its polarisation vector rotates by an angle which is proportional to the anomalous MDM~\cite{Baryshevsky:1979,Lyuboshits:1980,Kim:1983,Fomin:2017},
\eq%
\begin{equation}
\angle \( \vec{\mathds P}_{\rm i}
	     \vec{\mathds P}_{\rm f} \)
	  =
	  (1+	\gamma_\tau
	  	\, a_\tau)
	  \, \Theta,  
\label{eq:theta}
\end{equation}
where
$\vec{\mathds P}_{\rm i}$ and $\vec{\mathds P}_{\rm f}$
are the vectors of the initial and final polarisations of $\tau$ lepton (before and after the crystal),
$\gamma_\tau$ is the Lorentz factor of $\tau$ lepton ($\gamma_\tau\gg1$), and
$\Theta = L/R$ is the deflecting angle of the bent crystal,
$L$ and $R$ being the length and bending radius of the crystal.

Thus, to measure  the anomalous MDM of $\tau$ we need to have a beam of polarised $\tau$ leptons 
directed at the crystal, and to know the $\tau$ initial and final polarisations and energy.
As the polarised $\tau$ leptons are produced from the decay of $D_s$ mesons, the crystal should be placed at 
a certain distance $L_{\rm v}$ from the target to maximise the probability of $\tau$ production 
(see Appendix~\ref{app:Void}).

\begin{figure}[b]
\begin{center}
\includegraphics[width=0.7\textwidth]{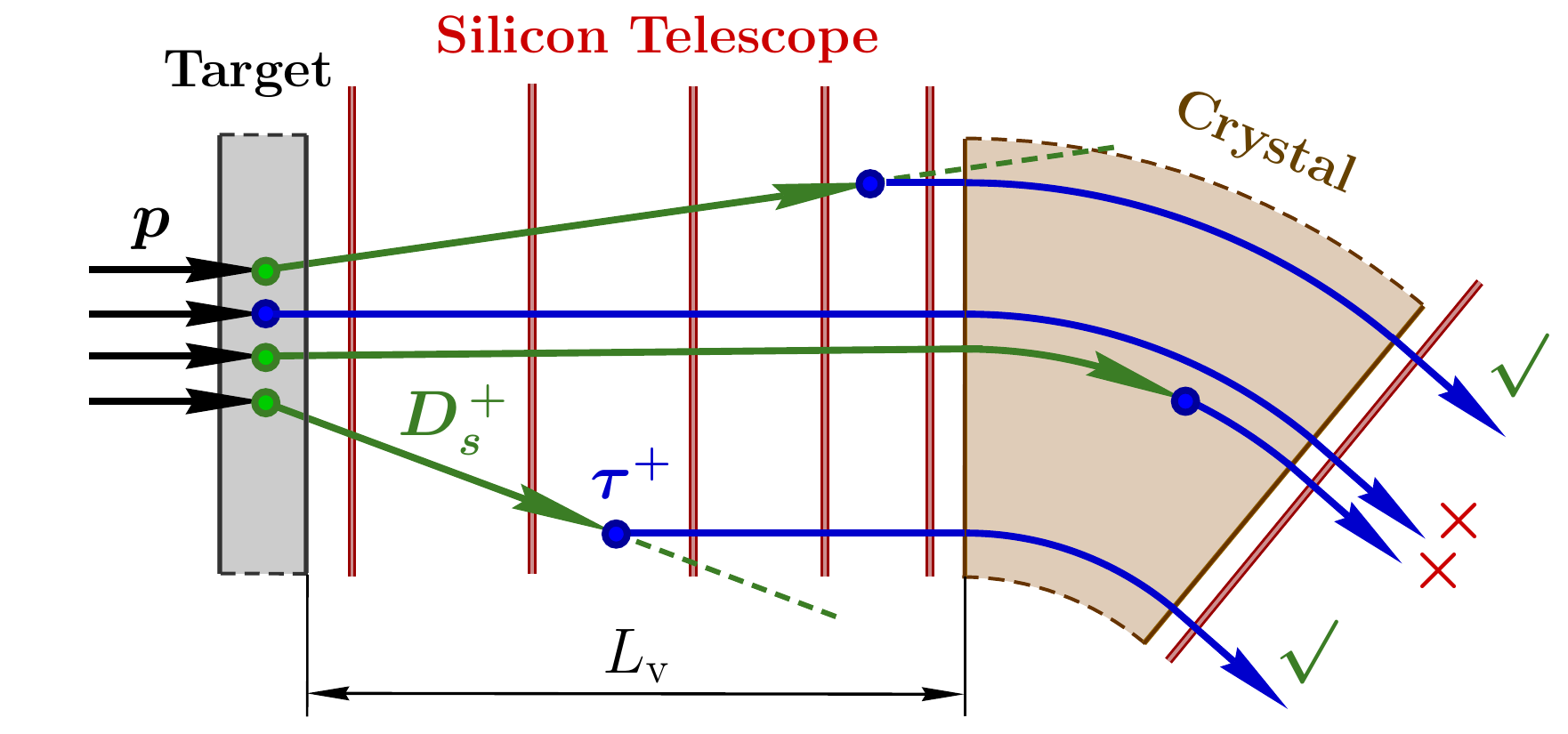}
\end{center}
\caption{Schematics of the setup with a silicon telescope detector.}
\label{fig:Sc_Tele}
\end{figure}

To reconstruct the direction of initial polarisation of Eq.~\eqref{eq:007}, we need to select $\t$ leptons coming from
$D_s$ decays and also to know the direction of the $D_s$ and $\t$ momenta and the energy of either 
$D_s$ or $\t$ particles.
For the present study we consider the $\tau$ decay into three charged pions,
so we need a suitable setup in order to get rid of the background processes leading to the same final state 
(see Sec.~\ref{subsec:Background}).
A solution could be to put a silicon telescope detector between the target and the crystal (see Fig.~\ref{fig:Sc_Tele}).
Further optimisation of the telescope design can be performed, however, the outlined 
structure is sufficient for a demonstration of the measurement principle, which is a purpose of the present article.

The first problem of this setup is that the telescope detector should have a very high angular resolution on a 
rather short base:
the distance between the target and the crystal should be $\sim\,10 - 40\,{\rm cm}$ (see Appendix~\ref{app:Void}),
and the angular resolution should be better than $\theta_{max}(E_D=2\,{\rm TeV})\approx100\,\mu{\rm rad}$.
The second problem is that the telescope detector in this case is very close to the target and could suffer from 
the background overload.

\begin{figure}[b]
\begin{center}
\includegraphics[width=0.8\textwidth]{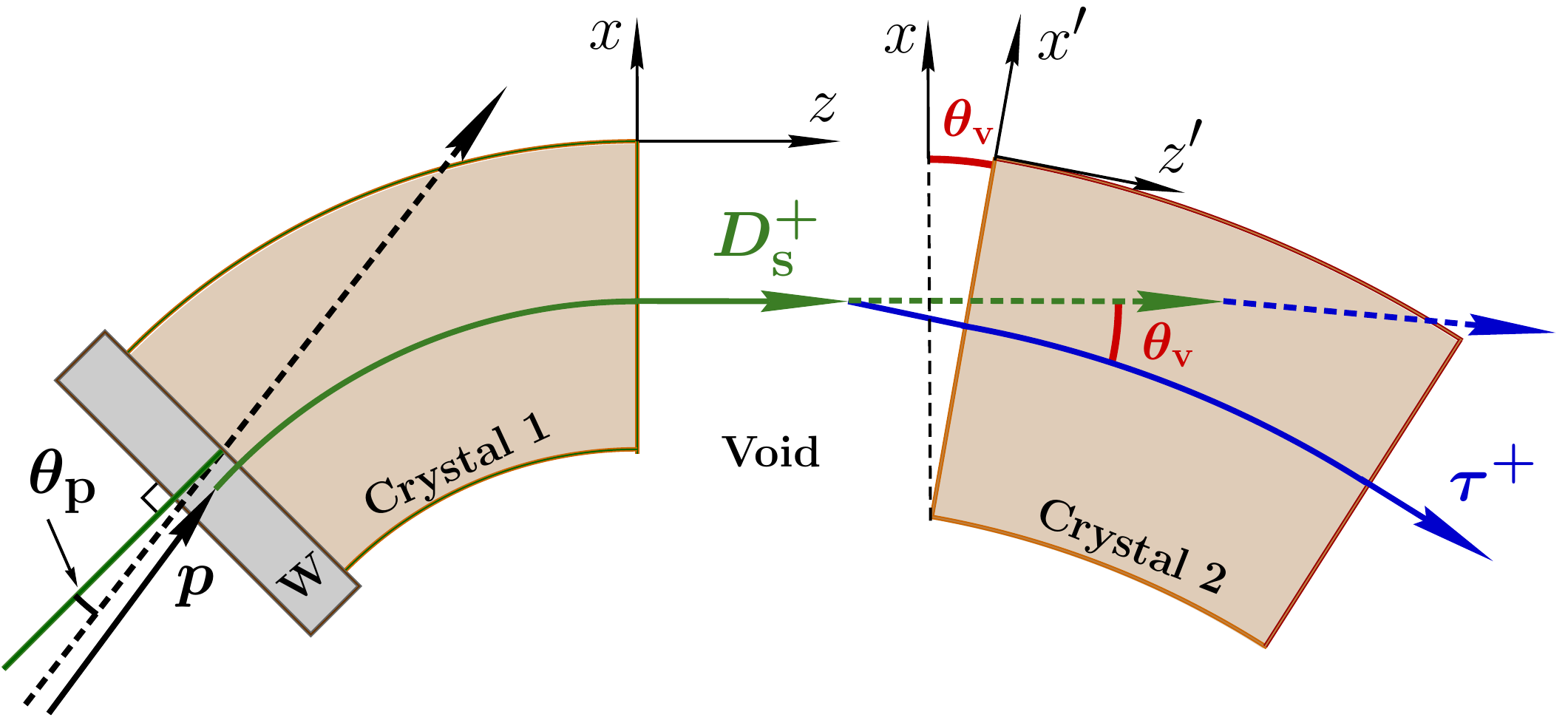}
\end{center}
\caption{Double crystal scheme.}
\label{fig:sc_Ds-T}
\end{figure}

In this concern, we propose a  two crystal setup as shown in Fig.~\ref{fig:sc_Ds-T}.
The purpose of the first crystal is to channel $D_s$ mesons and so to constrain their momentum direction. 
Misaligning the first crystal from the direction of the proton beam by a small angle $\q_p$ 
is to restrict the channeling of the initial protons.
As a result we would have a collimated beam of $D_s^+$ mesons satisfying the condition
$|\,\theta_x\,|<\theta_{\rm acc}\sim5\,\mu{\rm rad}$,
where
$\theta_x=\arcsin(p_{x}/p)$ and
$\theta_{\rm acc}$ is the angle of acceptance to the planar channeling regime
\cite{Fomin:2017,FominThesis}.

The second crystal is for channeling the $\tau$~lepton from the $D_s$ decay.
Misaligning the second crystal from the direction of the $D_s$ beam by an angle $\q_{\rm v}$ would select 
(by channeling) only $\tau$ leptons produced between the two crystals and would not permit the 
collimated $D_s$ beam to be channeled in the second crystal (see Fig.~\ref{fig:sc_Ds-T}).
The initial direction of the $\tau$ polarisation is reconstructed from the $D_s$ and $\tau$ momenta, which 
are selected by channeling in the first and the second crystal, respectively.

In the following, we will determine the optimal parameters of this setup: lengths and bending radii of two crystals, 
distance between crystals and orientation of the second crystal.
The direction of the final polarisation can be reconstructed by analysing the angular distribution of 
the $\tau$ decay products, as we discussed in Sec.~\ref{subsec:Final Polarisation}.
Finally, we obtain from \eqref{eq:theta}  the value of the anomalous MDM $a_\t$.

\newpage


\section{\label{sec:sensitivity} Sensitivity studies}

In this section we present a sensitivity study for measuring the MDM and EDM of the $\tau$ lepton using the setup described 
in the previous section.
The absolute statistical error of the measurement of the anomalous magnetic moment can be written in the 
following way,
\eq%
\begin{equation}
\Delta a_\t	   =	\sqrt{\dfrac{1}{
					\,N_{\rm POT}
					\ \eta_{\rm det} 
					\ Br
					\ S^2
					\ \eta_{\rm \,setup}\,
					}},
			\qquad
\eta_{\rm \,setup}   =	\Theta^2
				\displaystyle \int {\rm d}\e
				\,\frac{\pd N_\t^{\rm def}}{\pd\e}
				\ \g_\t^2\ 
				\,{\mathds P}^2,			
\label{eq:da}
\end{equation}
where $N_{\rm POT}$ is the total number of protons on target,
$\eta_{\rm det}$ is the detector efficiency,
$Br$ is the branching fraction of the selected $\tau$ decay,
$S$ is the sensitivity of polarisation reconstruction by the analysis of the $\tau$ decay,
$\eta_{\rm \,setup}$ is the setup efficiency (target plus two crystals),
$\pd N_\t^{\rm def}/d\e$ is the spectrum of deflected $\tau$ leptons normalised to one initial proton,
$\g_\t$ and ${\mathds P}^2={\mathds P}_x^2+{\mathds P}_z^2$ are the Lorentz factor and the square of 
averaged polarisation (in $xz$ plane) of the deflected $\tau$ lepton as a function of its energy.
In the following sections, we discuss these terms and give some estimates on their values.

The error of MDM connected with the precision of $\tau$ energy reconstruction \eqref{eq:dE} can be written in the following way,

\eq%
\begin{equation}
\Delta a^{(\gamma)}=\frac
{\overline a + \gamma^{-1}}
{1+\gamma/\Delta\gamma}\approx
\frac{\overline a_\tau+0.001}{6}.
\label{eq:Da}
\end{equation}
Considering the expected value of anomalous MDM, $\overline a_\tau \sim 0.001$ and Lorentz factor, $\gamma\sim1000$ this contribution is negligible.

Due to the MDM, the spin precession takes place only in the $xz$ plane. If $\tau$ lepton possesses the EDM, the $y$ component of polarisation will also be induced. 
In the limit $\gamma_\tau\gg1$, and assuming that EDM effects are small compared to the MDM spin precession, using formulas in~\cite{Bagli:2017}, the ratio between absolute error on EDM and MDM is
\eq%
\begin{equation}
\frac{\Delta f_\tau}{ \Delta a_\tau } \approx
\left| \frac{ 4\, \gamma\, \overline{a}_\tau}{\Theta\, (1+\gamma \overline{a}_\tau)^2} \right|
\label{eq:Dd/Da}
\end{equation}

Analogous to \eqref{eq:da} the absolute statistical error of the measurement of the EDM can be written in the following way,

\begin{equation}
\Delta f_\tau	   =	\sqrt{\dfrac{1}{
					\,N_{\rm POT}
					\ \eta_{\rm det}
					\ Br
					\ S^2 
					\ \eta^{(f)}_{\rm \,setup}\,
					}},
			\qquad
\eta^{(f)}_{\rm \,setup}   =	\frac{\Theta^4}{\,16\, \overline a_\t^{\,2}\,}
				\displaystyle \int {\rm d}\e
				\,\frac{\pd N_\t^{\rm def}}{\pd\e}
				\ (1+\g_\t\, \overline a_\t)^4\ 
				\,{\mathds P}^2,			
\label{eq:dd}
\end{equation}


\subsection{\label{subsec:Ds} First crystal: spectrum of deflected $D_s$ mesons}

\begin{figure}[b]
\begin{center}
\includegraphics[width=.92\textwidth]{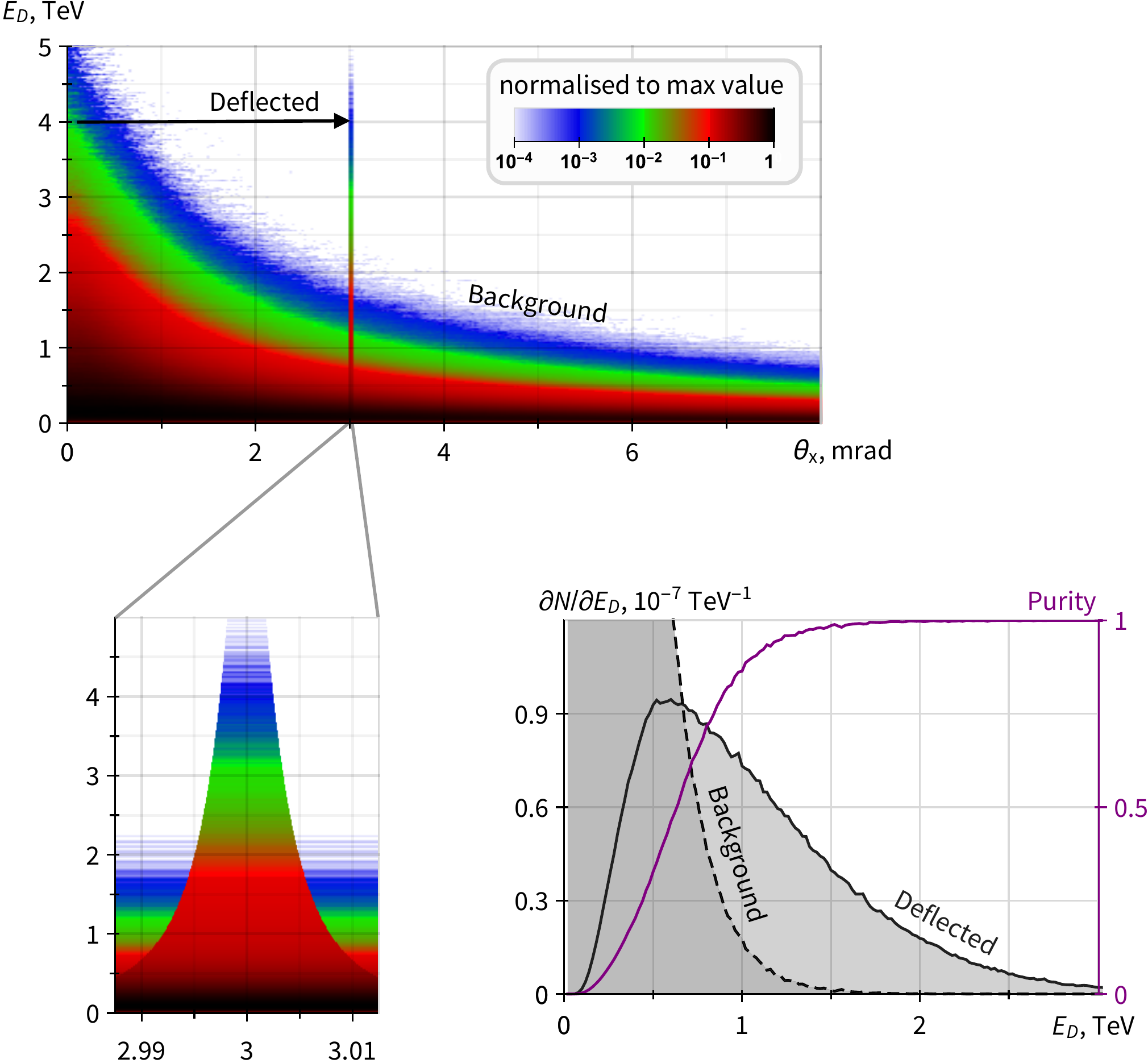}
\end{center}
\caption{Spectral angular distribution of $D_s$ mesons after the first crystal. 
``Deflected'' describes a collimated $D_s$ beam deflected by a germanium crystal by $\Theta_D=3\,$mrad;
``Background'', $D_s^+$ produced in the target.
Purity is the ratio of deflected $D_s^+$ to all $D_s^+$ mesons (initially produced or deflected) at $\q_x$.
$\q_x=0$ corresponds to the direction of proton beam.
}
\label{fig:Ds_dEdQ}
\end{figure}

The spectral angular distribution of $D_s$ mesons produced by $7\,{\rm TeV}$ protons on a fixed target 
 is obtained simulating 
minimum-bias $pp$ collisions with the Pythia~8.2 generator~\cite{Sjostrand:2007gs} and shown in 
Fig.~\ref{fig:Ds_dEdQ}.
In the current study we choose a tungsten target with depth $L_{\rm tar}=1\,{\rm cm}$ 
along the direction of the incident proton beam. As it was shown in Ref.~\cite{Fomin:2017}, the target depth is 
limited only by the increasing number of secondary particles produced in deeper targets.

Assuming no nuclear effects the production rate of $D_s^+$ mesons per one incident proton is
\eq%
\begin{equation}
N_D^{\rm\,tar}=
 	\rho
  	\,N_{\rm{A}} 
   	\,\sigma_D \,
    	\frac{A_{\rm tar}}{M_{\rm tar}}
    	\,L_{\rm tar},
 \label{eq:production}
 \end{equation}
where
$\rho$, $A_{\rm tar}$ and $M_{\rm tar}$ are the density,  number of nucleons and molar mass of the target, respectively,
$N_{\rm A}$ is the Avogadro number,
and
$\sigma_D$ is the cross section of $D_s^+$ production in $pp$ collision at center-of-mass energy
$\sqrt{s}=115\,$GeV (corresponding to the fixed target experiment at the $7\,$TeV LHC proton beam).
For the chosen target, $N_D^{\rm\,tar}\approx1.4\times10^{-4}$.

In the setup shown in Fig.~\ref{fig:sc_Ds-T}, we propose to place a crystal with the purpose of preparing a 
collimated energetic beam of $D_s$ mesons.
The initial direction of the crystallographic plane is chosen as $\q_p=100\,\mu$rad.
Thus, after the first crystal, we can have two categories of $D_s$ meson beams.
The first one, referred from now on as ``Background'', is composed of $D_s^+$ mesons which are initially 
produced at an angle $\q_x\sim\Theta_D$. In this case, the $D_s$ mesons are poorly collimated, and as 
a consequence, the average $\tau$ polarisation is close to zero.
The second category, further referred as ``Deflected'', is composed of $D_s^+$ mesons deflected by the crystal by an angle $\Theta_D$.
These $D_s$ mesons are collimated and hence can be the source of polarised $\tau$ leptons.

The purity, which is defined as the ratio between the number of deflected $D_s$ mesons and the total (``Background''+``Deflected'') number of $D_s$ mesons, is increasing as a function of energy as can be seen on Fig.~\ref{fig:Ds_dEdQ}.
For the sensitivity analysis, we take into account this effect by introducing an effective spectra, which is obtained by convolution of the spectra with the purity.
The background from the $D_s$ produced at $\Theta_D$ is specifically considered, while a preliminary consideration of other sources is addressed in section~\ref{subsec:Background}.

The spectral angular distribution of $D_s$ mesons deflected by a bent crystal (Fig.~\ref{fig:Ds_dEdQ}, Deflected) is obtained using a parameterisation based on the Monte Carlo simulation of particle propagation through the crystal~\cite{FominThesis}, accounting the incoherent scattering on the electron subsystem of a crystal and on thermal vibrations of the atoms at lattice nodes, that leads to dechanneling effect.
The optimal bending radii and lengths were found for a silicon crystal to be $R_D=15\,$m and $L_D=4.5\,$cm, and for a  germanium crystal, $R_D=10\,$m and $L_D=3\,$cm.
Thus, the optimal deflection angle is $\Theta_D\sim3\,{\rm mrad}$.
From our simulation, we determined that the $D_s$ deflection rate per one incident proton on target by a germanium crystal is about  $1.1\times10^{-7}$ and about $0.6\times10^{-7}$ with a silicon crystal.


\subsection{\label{subsec:Tau} Second crystal: spectrum and polarisation of deflected $\tau$ leptons}

For an optimal rate of $\tau$ leptons from $D_s$ decays, the second crystal should be placed at a distance 
$L_{\rm v}\sim10\,{\rm cm}$ from the first one (see Appendix~\ref{app:Void}).
This way, about $1.3\,\%$ of $D_s$ mesons will decay into $\tau$ leptons before the entrance of the second 
crystal.

\begin{figure}[b]
\begin{center}
\includegraphics[width=.95\textwidth]{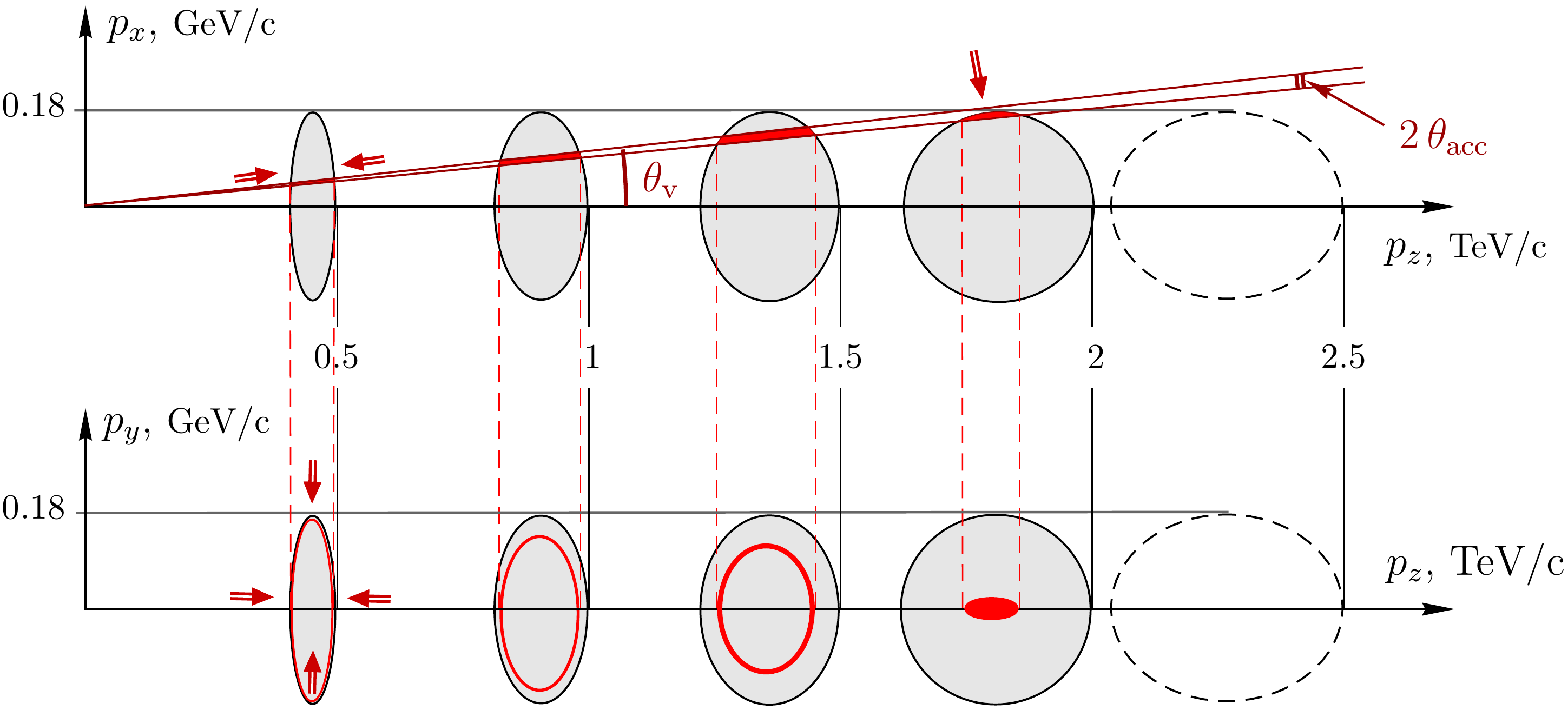}
\end{center}
\caption{Collimation of $\tau$ momenta by the second crystal.
$p_z$ is chosen in the direction of  the $D_s$ momentum.
Each ellipsoid represents the phase space of $\t$ leptons produced by $D_s$ mesons of specific energy.
The red area shows the phase space of $\t$ leptons collimated by the second crystal.
Arrows show the polarisation of $\t$. Scale on $p_x$ is $\sim10^4$ times smaller than on $p_z$ because $\gamma_D\gg1$.}
\label{fig:Sc_Coll_pxpy}
\end{figure}

The momentum distribution of the $\t$ leptons produced in the two-body decay is defined by the momentum of 
the $D_s$ mesons. In the laboratory frame, the $\tau$ momenta form an ellipsoid of rotation.
Figure~\ref{fig:Sc_Coll_pxpy} represents the $\t$ momentum distribution in the $p_xp_z$ and $p_yp_z$ 
phase spaces in the laboratory frame.
The two red lines at the angles $\q_{\rm v}\pm\q_{\rm acc}$ from the $p_z$-axis highlight the area of phase 
space of $\tau$ momentum that can be deflected by the second crystal.
In $p_y\,p_z$ phase space the deflected area has an elliptical shape which results from intersection of the second crystal plane with the ellipsoid of the $\tau$ momenta.

As can be seen from the figure, the misalignment of the two crystals leads to a large suppression of the 
spectra of deflected $\t$ leptons.
For instance, the angle between the crystals $\theta_{\rm v}=100\,\mu$rad (see Fig.~\ref{fig:sc_Ds-T}) matches the
maximal allowed angle between the $\t$ and $D_s$ momenta 
given in Eq.~\eqref{eq:010} for a $D_s$ energy of $E_D=2\,$TeV,
thus, $\t$ leptons produced by more energetic $D_s$ mesons would not be deflected.
Nevertheless, the deflected $\t$ leptons produced by $2\,$TeV $D_s$ mesons would be almost 
$100\,\%$ polarised in the $x$ direction, unlike those produced by less energetic $D_s$ mesons as can be
seen in Fig.~\ref{fig:Sc_Coll_pxpy}.

The spectra of deflected $\t$ leptons can be described by the following expression
\eq%
\begin{equation}
\fr{\pd N_\t^{\rm def}}{\pd\e_\t}   =	\int\limits_0^{\e_{\rm max}} {\rm d} E_D
					\,\fr{\pd N_D}{\pd E_D}
					\ \eta\,(L_{\rm v}, E_D)
					\ \fr{\pd N_\t}{\pd\e_\t} ( E_D)
					\ \eta_{\,\rm coll} \(  E_D, \e_\t \)
					\ \eta_{\,\rm chan}\,(\e_\t),
\label{eq:deflected}
\end{equation}
where
$\pd N_D/\pd E_D$ is the effective spectra of deflected $D_s$ mesons (accounting for the purity), 
$\eta(L_{\rm v}, E_D)$ is the conversion efficiency of $D_s^+$ into $\tau^+$ given in Eq.~\eqref{eq:Conversion},
$\pd N_\t/\pd\e_\t ( E_D)$ is the spectra of $\tau$ leptons produced by $D_s$ decays as a function of 
the $D_s$ energy, $\eta_{\,\rm chan}(\e_\t)$ is the channeling efficiency~\cite{FominThesis} taking into account 
the $\tau$ decay, $\eta_{\,\rm coll}( E_D, \e_\t) = \( \phi_{\rm max}-\phi_{\rm min} \) / \pi $ is the fraction of $\t$ 
leptons collimated when entering the crystal, with
\eq%
\begin{equation}
\sin \phi			=	\fr{\q_x}{\q( E_D,\e_\t)}, \quad
\sin \phi_{\rm max}	=	\fr{\q_v \, + \q_{\rm acc}(\e_\t)}{\q( E_D,\e_\t)}, \quad
\sin \phi_{\rm min}	=	\fr{\q_v \,  - \q_{\rm acc}(\e_\t)}{\q( E_D,\e_\t)}.
\end{equation}
\eq%

The expressions for the projections of polarisation on the $x$ and $z$ axes in the relativistic case are
\eq%
\begin{equation}
{\mathds P}_x( E_D, \e_\t, \phi)	=		{\mathds P}_\T( E_D, \e_\t) \, \sin \phi, \qquad
{\mathds P}_z( E_D, \e_\t)		\approx	{\mathds P}_\ll( E_D, \e_\t).
\label{eq:Pxz}
\end{equation}

By averaging over the $D_s$ spectra and taking into account the collimation due to channeling, we get the 
expression of the $x$ and $z$ components of polarisation of deflected $\tau$ leptons as a function of energy,
\eq%
\begin{equation}
{\mathds P}_z(\e_\t)	=	\fr{1}{\pd N_\t^{\rm def}/\pd \e_\t}
			\int {\rm d} E_D
			\ \fr{\pd N_D}{\pd  E_D}
			\ \fr{\pd N_\t^{\rm def}}{\pd \e_\t} ( E_D)
			\ {\mathds P}_z ( E_D, \e_\t),
\label{eq:Pz(e_t)}
\end{equation}

\begin{equation}
{\mathds P}_x(\e_\t)	=	\fr{1}{\pd N_\t^{\rm def}/\pd \e_\t}
			\int {\rm d} E_D
			\,\fr{\pd N_D}{\pd  E_D}
			\,\fr{\pd N_\t}{\pd \e_\t} ( E_D)
			\,{\mathds P}_\T ( E_D, \e_\t)
			\,\fr{1}{\pi}  \int\limits_{\phi_{min}(\e_\t)}^{\phi_{max}(\e_\t)}
			\,\sin\phi \ {\rm d}\phi.
\label{eq:Px(e_t)}
\end{equation}
The integration over $\phi$ in Eq.~\eqref{eq:Px(e_t)} means averaging over the direction of particles in crystal plane and over acceptance angle in the direction perpendicular to this plane.
In Eq.~\eqref{eq:Pz(e_t)} this integration is trivial, and is simply the term $\eta_{\rm coll}$ from Eq.~\eqref{eq:deflected}.
Due to the symmetry of the setup the average of $y$-component of the initial polarisation is zero.

\begin{figure}[t]

\begin{center}
\includegraphics[width=0.49\textwidth]{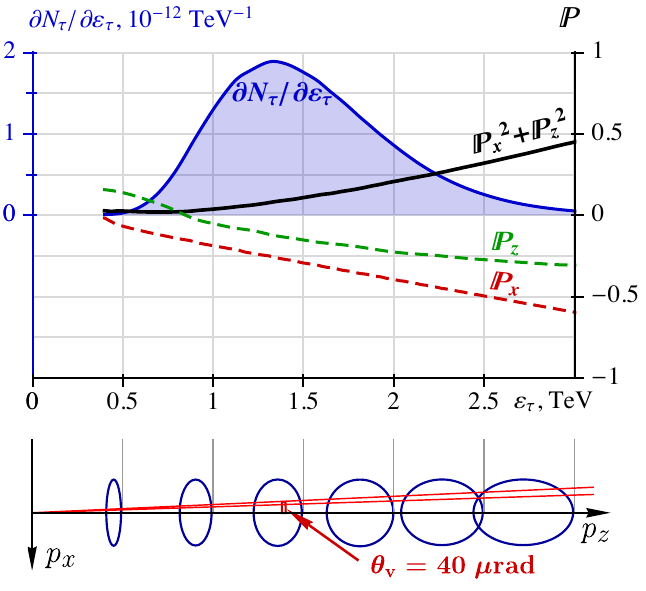}\hfill
\includegraphics[width=0.49\textwidth]{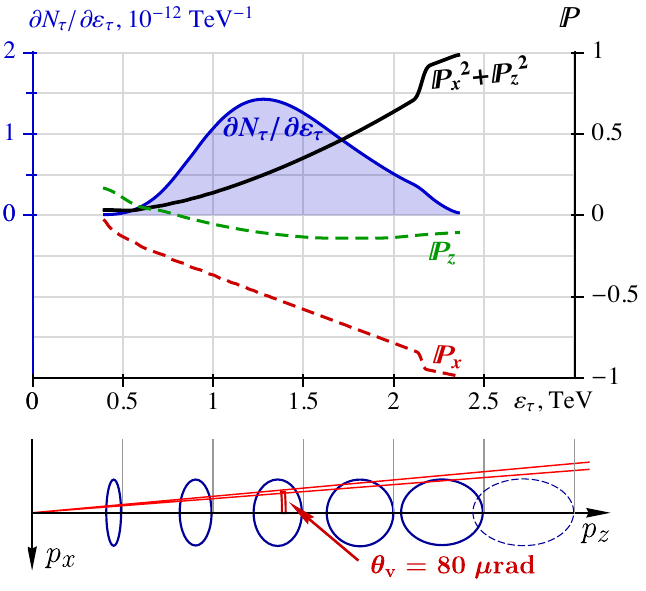}\\
\end{center}
\caption{
Polarisation and spectra of deflected $\tau$ leptons
for (left) $\q_{\rm v}=40\,\mu{\rm rad}$ and (right) $\q_{\rm v}=80\,\mu{\rm rad}$.
The spectra are normalised to one initial proton. The red and green dashed lines represents the
 $x$ and $z$ components of the averaged $\tau$ polarisation.
}
\label{fig:Ds-T_PE}
\end{figure}

In Fig.~\ref{fig:Ds-T_PE} we present the spectra given in Eq.~\eqref{eq:deflected} and the polarisation as a function 
of energy given in Eqs.~\eqref{eq:Pz(e_t)} and \eqref{eq:Px(e_t)} of deflected by 12 cm germanium crystal $\tau$ leptons for two orientations of 
the second crystal, (left) $\q_{\rm v}=40\,\mu{\rm rad}$, $R=9\,$m, and (right) $\q_{\rm v}=80\,\mu{\rm rad}$, $R=7\,$m.

In the second case, the number of deflected $\tau$ leptons is smaller and the spectra is softer, but the degree of polarisation is significantly greater.

\begin{figure}[b]

\begin{center}
\includegraphics[width=.49\textwidth]{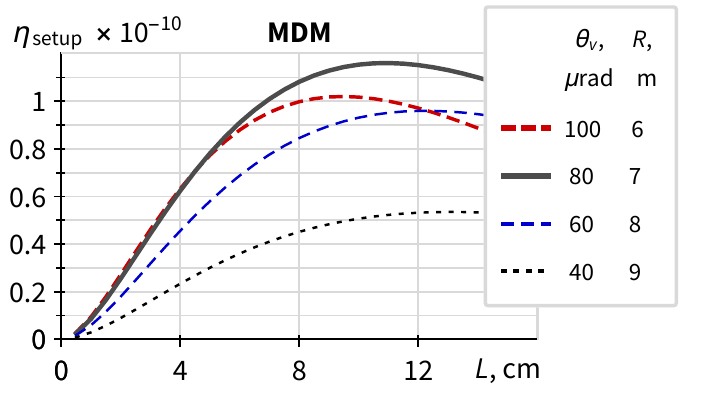}
\includegraphics[width=.49\textwidth]{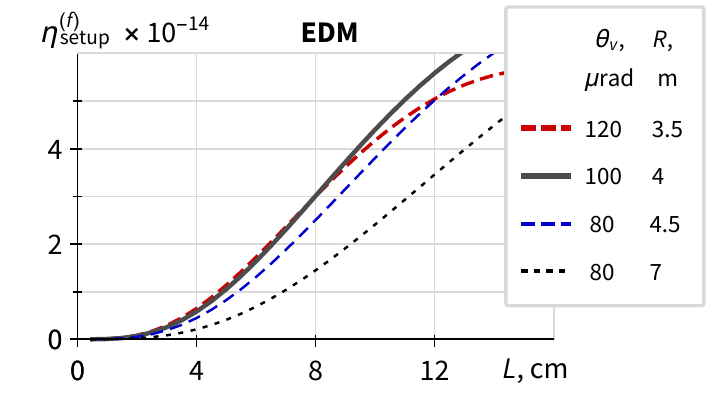}
\end{center}
\caption{
Setup efficiencies (left) for MDM measurement $\eta_{\rm setup}$ and (right) for EDM measurement $\eta^{(f)}_{\rm \,setup}$ as a function of the second crystal length, $L$, for various values of 
the angle $\q_{\rm v}$ and optimal radii of curvature $R$ (listed on the right).
}
\label{fig:N_Lcr2}

\end{figure}

By varying the orientation and dimensions of the second crystal, we find the maximal setup efficiency for MDM measurement, 
$\eta_{\rm setup}\sim1.1\times10^{-10}$ per incident proton (see Fig.~\ref{fig:N_Lcr2}, left).
For this case, the second crystal parameters are the following: material --- germanium, orientation angle $\q_{\rm v}=80\,\mu{\rm rad}$, 
length $L=10\,{\rm cm}$ and bending radius $R=7\,{\rm m}$ and deflection angle 
$\Theta\approx14\, {\rm mrad}$.
If the crystal length is limited to $6\,{\rm cm}$, it is better to increase the orientation angle to 
$\q_{\rm v}=100\,\mu{\rm rad}$, so that the optimal bending radius will decrease to $R=6\,{\rm m}$, and hence 
the deflection angle would be $\Theta=10\,{\rm mrad}$. In such case, the setup efficiency would be 
$\eta_{\rm setup}\sim0.9\times10^{-10}$.

In the EDM measurement, comparing to the MDM measurement, the deflection angle is more important than Lorentz factor and the number of deflected particles (see Eq.~\eqref{eq:Dd/Da}).
For this reason the optimal bending radius is smaller (see Fig.~\ref{fig:N_Lcr2}, right) and the optimal crystal length is greater.
Dotted line in Fig.~\ref{fig:N_Lcr2} (right) shows the EDM measurement efficiency when the crystal parameters are optimised for the MDM measurement.
Optimising the setup for EDM measurement can gain a factor of two to the efficiency.


\subsection{\label{subsec:Background} Preliminary study of the background from other decay channels}

In the current study we consider the $\tau$ decay into three charged pions,
\eq%
\begin{equation}
\t^+\to\pi^+\pi^+\pi^-\,\overline{\nu}_\tau.
\label{eq:decay}
\end{equation}
The advantage of this decay channel is that the secondary vertex can be reconstructed and the final $\tau$ polarisation
 can be mainly reconstructed from the momenta of pions, which is also very important to maximise the value of $S$.
We now consider possible processes producing $\pi^+\pi^+\pi^-$ secondary vertices at the end of the second crystal.

\begin{figure}[t]
\begin{center}
\includegraphics[width=0.99\textwidth]{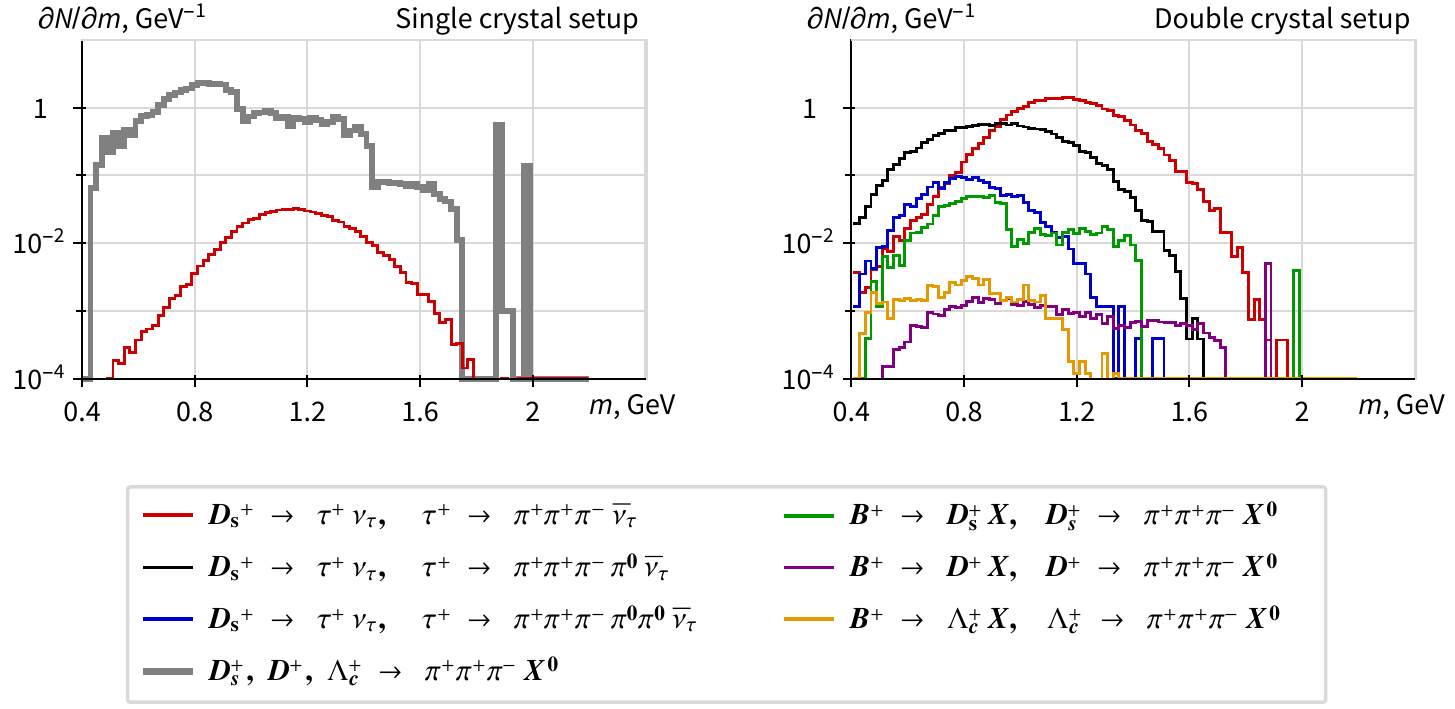}
\end{center}
\caption{Invariant mass distribution of $\pi^+\pi^+\pi^-$ combinations from (red line) the main decay chain,
$D_s^+\to\t^+\,\nu_\t,\ 
\t^+\to\pi^+\pi^+\pi^-\,\overline{\nu}_\t$,
and for the background decay chains with $\pi^+\pi^+\pi^-$ in the final state and listed at the bottom,
using (left) a single-crystal setup and (right) a double-crystal setup.}
\label{fig:dm_Background}
\end{figure}

The double-crystal setup plays an important role for the background separation.
Due to a small misalignment of the second crystal $\q_{\rm v}\sim100\,\mu{\rm rad}$, the particle deflected by the 
first crystal cannot be involved into the channeling regime in the second one.
This removes all charged $D$ mesons decaying directly into $\pi^+\pi^+\pi^-$, as shown in 
Fig.~\ref{fig:dm_Background}.
In addition, the planar channeling in a long crystal is possible only for positively charged particles.
As a consequence, $\pi^+\pi^+\pi^-$ events can only come from the chain
\mbox{$pp\to X^+...\to Y^+...\to\pi^+\pi^+\pi^-...\,$}.

To study the background from these decay channels, we used the Pythia~8.2 generator.
Figure~\ref{fig:dm_Background} shows the distribution over the invariant mass of the three reconstructed pions for 
the main decay channels, listed on the figures.
The simulation shows that the main source of background is the decay
$\tau^+\to\pi^+\pi^+\pi^-\pi^0\,\overline{\nu}_\t$.
With a requirement on the invariant mass $\ge1\,{\rm GeV}$, we can reduce its contamination to $25\,\%$.
For the present study, we suppose that the $\tau$ polarisation from the $\pi^+\pi^+\pi^-\pi^0$ decay channel is zero.
This is a pessimistic assumption since the presence of $\pi^0$ should not completely break the correlation between 
the $\tau$ polarisation and the angular distribution of $\pi^+\pi^+\pi^-$.
In addition, this background could be further reduced by the identification of the $\pi^0$ meson.
Other sources of background come from $B$-meson decays, but as shown in Fig.~\ref{fig:dm_Background}, they 
are negligible.
A detailed study could be performed to evaluate the background coming from accidental reconstruction of three 
pion vertices.


\subsection{\label{subsec:Results} Results of the sensitivity studies}

In this section, we study the efficiency of the double-crystal setup proposed in the previous section and 
determine the optimal parameters of the setup for a $7\,{\rm TeV}$ proton beam. They are:
\begin{itemize}
\item Target: tungsten with $L_{\rm tar}=1\,$cm;
\item Crystal 1: germanium with  $L_{D}=3\,{\rm cm}$, $R_{D}=10\,{\rm m}$ and $\q_{p}=100\,\mu{\rm rad}$;
\item Crystal 2: germanium with $L=10\,{\rm cm}$, $R=7\,{\rm m}$ and $\q_{\rm v}=80\,\mu{\rm rad}$, at $L_{\rm v}=10\,$cm.
\end{itemize}
\begin{table}[!t]
\caption{Efficiency of the measurement: current (by process) and total (per incident proton).}
\begin{center}
\begin{tabular}{ |l |l |c | c | c | c|}
\hline
~~~~~~~~& ~~~~~~~~~~~&\multicolumn{2}{c|}{double crystal setup} &
\multicolumn{2}{c|}{single crystal setup} \\
~Place~~~~~~~& ~process (factors)~~~~~~~~~~&~~~current~~~ & ~~~~~total~~~~~ & ~~~current~~~ &~~~~~~total~~~~~~\\
\hline
~target &~$~p\to D_s^+$,~~$D_s^+$ decay&$1.1\x10^{-4}$ & $1.1\x10^{-4}$ & $1.1\x10^{-4}$ & $1.1\x10^{-4}$ \\  
~crystal 1 &$~D_s^+$ collimation&         $0.8\x10^{-2}$ & $0.9\x10^{-6}$   &  $-$ & $-$ \\  
~crystal 1 &$~D_s^+$ deflection&           $1.3\x10^{-1}$ & $1.1\x10^{-7}$   &  $-$ & $-$ \\  
~void &~$D_s^+\to\tau^+$&                     $1.2\x10^{-2}$ & $1.4\x10^{-9}$   & $0.5\x10^{-3}$ & $0.5\x10^{-7}$ \\  
~crystal 2 &~$\tau^+$ collimation&          $0.5\x10^{-1}$ & $~0.7\x10^{-10}$& $1.7\x10^{-2}$ & $0.9\x10^{-9}$ \\  
~crystal 2 &~$\tau^+$ deflection &          $0.3\x10^{-1}$ & $~2.2\x10^{-12}$ & $0.3\x10^{-1}$ & $~0.3\x10^{-12}$ \\  
~detector &~$\eta_{\rm det}\times Br$&  $0.4\x10^{-1}$ & $~1.0\x10^{-13}$ & $0.4\x10^{-1}$ &  $~1.3\x10^{-12}$ \\  
~reconstr. &~$\eta_{\rm \,MDM}\sim\langle{\mathds P}^2\,\gamma^2\rangle\,\theta^2\,S^2$&
                                                                $\sim\,$8&  $\sim0.7\x10^{-12}$    & $\sim\,$0.5 & $\sim0.7\x10^{-12}$ \\  
~reconstr. &~$\eta_{\rm \,EDM}\sim\langle{\mathds P}^2\,\gamma^2\rangle\,\theta^4\,S^2$&
                                                                 $\sim\,$$1.5\x10^{-3}$ &  $\sim\,$$1.5\x10^{-16}$   & $\sim1.1\x10^{-4}$ & $\sim1.5\x10^{-16}$ \\  
\hline
\end{tabular}
\end{center}
\label{tab:Efficiency}
\end{table}
In this study, we did not look in details at the detection system.
We consider the analysis of the decay channel $\tau^+\to\pi^+\pi^+\pi^-\overline{\nu}_\tau$ ($Br=0.0899$~\cite{PDG:2018})
and suppose a detection efficiency of $\eta_{\rm det}=50\,\%$.
In Table~\ref{tab:Efficiency} we present (column~3) the efficiencies of each process, starting from $D_s^+$ production in the target, up to reconstruction of MDM and EDM of $\tau$ lepton, and (column~4) the total efficiencies, i.e. the number of $D_s^+$ or $\tau$ after each stage of double crystal setup referred to the total number of impinging protons on target.
Last two rows in the table present efficiencies of MDM and EDM reconstruction.
Using latter, we can rewrite Eqs.~\eqref{eq:da} and~\eqref{eq:dd} in a following way
\eq%
\begin{equation}
\Delta a_\tau=\frac{1}{\sqrt{N_{\rm POT}\,\eta_{\rm MDM}}}\qquad\qquad
\Delta d_\tau=\frac{\mu_{\rm B}}{2\,\sqrt{N_{\rm POT}\,\eta_{\rm EDM}}}
\label{eq:sigmas}
\end{equation}
where $\Delta d_\tau$ is the absolute statistical error of the measured EDM in units [e cm] and $\mu_{\rm B}$ is the Bohr magneton.

We also estimate the efficiency of the single crystal setup presented in Fig.~\ref{fig:Sc_Tele} (see columns~5 and 6 in Table~\ref{tab:Efficiency}).
The numbers in the table correspond to the following properties of the single crystal setup: germanium crystal with $L=10\,{\rm cm}$, $R=7\,{\rm m}$ and $\q_{\rm p}=100\,\mu{\rm rad}$ at $L_{\rm v}=40\,$cm, from $1\,$cm tungsten target.
We found that in order to reach the reconstruction efficiency of double crystal setup the conversion length should not be 
greater than $L_{\rm v}=40\,$cm.
On the other hand, such length is too short to reconstruct the $D_s^+ \to \tau^+ \,\nu_{\tau}$ vertex, i.e. to verify that the signal comes from the $\tau$ lepton produced before the crystal, but not inside or after the one, or from other particle with a similar final state (see Fig.~\ref{fig:dm_Background}).
Due to these reasons we consider the double crystal setup to be more efficient. 
Note, that due to a greater average polarisation, the double crystal setup has a better reconstruction efficiency of MDM and EDM (referred to one deflected $\tau$ lepton).

\begin{figure}[t]
\begin{center}
\includegraphics[width=.49\textwidth]{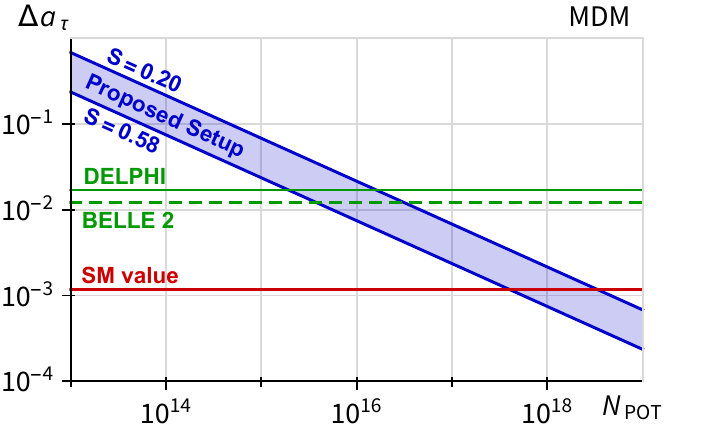}
\includegraphics[width=.49\textwidth]{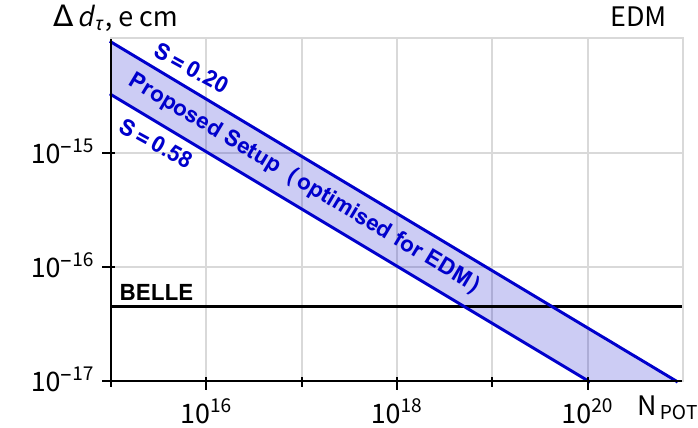}
\end{center}
\caption{Absolute statistical error (left) of the anomalous MDM, (right) of EDM of the $\tau$ lepton.}
\label{fig:Result}
\end{figure}

Figure~\ref{fig:Result} (left) shows the absolute statistical error of the measured anomalous MDM of the $\tau$
lepton using Eq.~\eqref{eq:da} as a function of the total number of protons on target.
The blue lines show the sensitivity of the present analysis by varying in particular the value of $S$
in the range $0.2<S<0.58$.
The maximal value corresponds to the limit on the accuracy, in case all observables are fully reconstructed.
The green lines show the limits obtained by the DELPHI collaboration~\cite{Abdallah:2003} and expected for the BELLE$\ 2$ experiment~\cite{Eidelman:2016}.
The red line represents the value of the $\tau$ anomalous MDM predicted by the SM~\cite{Eidelman:2007}.
Further detailed studies will be performed to improve the current efficiency.
In particular, detailed background studies could allow us to use one prong $\tau$ decays. 

For EDM measurement this setup is not very efficient. Even when optimised for EDM measurement, the current experimental accuracy on EDM~\cite{Inami:2002} could be achieved only after about $10^{19}\,$POT, which is probably not realistic (see Fig.~\ref{fig:Result} (right)).


\section{\label{sec:Conclusions} Conclusions}

In this paper, we propose a method of a direct measurement of the anomalous magnetic dipole moment 
(MDM) of the $\tau$ lepton by studying its spin precession using a double-crystal setup at the LHC.
Our study shows that with this proposal, one can reach the present accuracy on the $\tau$ MDM using 
$10^{16}$ protons on target.
A total of $10^{18}$ protons on target are needed to reach an accuracy equivalent to the Standard Model value.
The measurement of EDM is even more challenging and a further order of magnitude of POT is needed to reach the present experimental accuracy.
The benefit of this method is that there is no assumption on the value of the $\tau$ electric dipole moment.
This study opens up the possibility of doing a single dedicated experiment to measure both the MDM of charmed 
baryons~\cite{Baryshevsky:2016,Fomin:2017,Botella:2016,Bagli:2017,Baryshevsky:2017,FominThesis}
and of $\tau$ lepton at the LHC.



\appendix
\section{\label{app:1} Polarisation of $\tau$ leptons in $D_s$ decays}

We consider the decay of the $D_s$ meson 
\eq%
\begin{equation}
D_s^+ (p) \to \tau^+ (k) +\nu_\tau (q),
\label{eq:a1} 
\end{equation}
with $p =k +q$, and find the polarisation vector of the $\tau$ lepton. The four-momentum of the $D_s$ meson is 
$p^\mu = (E_D, \, \vec{p}_D)$. We further assume that the neutrino is massless.

To describe the polarisation of a fermion in its rest frame, one introduces the vector $\vec{\xi}$ (${|\vec{\xi}|\le1}$). 
In the frame in which fermion moves with the four-momentum $k^\mu = (\varepsilon_\tau, \, \vec{k}_\tau)$, 
the polarisation four-vector is~\cite{Beresteckii:1982}
\eq%
\begin{equation}
a^\mu = \left( \frac{\vec{k}_\tau \, \vec{\xi}}{m_\tau}, \, \, \vec{\xi} 
+ \frac{\vec{k}_\tau \, ( \vec{k}_\tau \, \vec{\xi})}{m_\tau (m_\tau + \varepsilon_\tau)} \right), 
\label{eq:a2}
\end{equation}  
\eq%
with the conditions $k \cdot a =0$ and $a^2 = - \vec{\xi}^2$. 

We can write matrix element of the $D_s^+$ decay as 
\eq%
\begin{equation}
{\cal M} = \frac{g^2}{ 8 M_W^2} f_{D} \,V^*_{cs} \, p^\mu \, \bar{u}(q) \gamma_\mu (1- \gamma_5) v (k),
\label{eq:a3}
\end{equation}
\eq%
where $f_{D}$ is the weak-decay constant of the $D_s^+ \to \tau^+ \nu_\tau$ decay, 
$M_W$ is the mass of the $W$ boson, $V_{cs}$ is the element of the Cabibbo-Kobayashi-Maskawa matrix,  
$g = e/ \sin \theta_W$, $e$ is the positron charge and $\theta_W$ is the weak mixing angle.

Using the spin-density matrix of the $\tau^+$  lepton,
\eq%
\begin{equation}
v(k) \bar{v} (k) = \frac{1}{2}(\vslash{k} - m_\tau) \, (1+ \gamma_5 \vslash{a}),
\label{eq:a4}
\end{equation}
one obtains the probability to find the $\tau^+$ lepton with polarisation $a^\mu$. It is proportional to  
\eq%
\begin{equation}
|{ \cal M}|^2 = \frac{g^4  m_\tau^2}{16 M_W^4} |V_{cs}|^2 f^2_{D} \,   
(k \cdot q  - m_\tau \, {p \cdot a}),
\label{eq:a5}
\end{equation}
\eq%
with $k \cdot q = (M_D^2 -m_\tau^2)/2$. 

Equation~(\ref{eq:a5}) allows one to find the $\tau$ polarisation vector $\vec{\mathds P}$ which arises in the decay process. For this we write   
\eq%
\begin{equation}
(k \cdot q  - m_\tau \, {p \cdot a})  \equiv k \cdot q  \, \left(1 + \vec{\xi} \cdot \vec{\mathds P}\right),
\label{eq:a6} 
\end{equation}
\eq%
\begin{equation}
\vec{\mathds P} = \frac{1}{k \cdot q } \left[ m_\tau \vec{p}_D \, - \, E_D \vec{k}_\tau \, 
+ \, \frac{\vec{k}_\tau \, (\vec{k}_\tau\cdot \vec{p}_D)}{m_\tau+\varepsilon_\tau}    \right]. 
\label{eq:a7}
\end{equation}
\eq%

The corresponding polarisation four-vector in an arbitrary frame has a form  similar to Eq.~(\ref{eq:a2}), namely 
\eq%
\begin{equation}
S^\mu = \left( \frac{\vec{k}_\tau \cdot \vec{\mathds P}}{m_\tau}, \, \, \vec{\mathds P} 
+ \frac{\vec{k}_\tau \, ( \vec{k}_\tau \cdot \vec{\mathds P})}{m_\tau (m_\tau + \varepsilon_\tau)} \right), 
\label{eq:a8}
\end{equation}  
\eq%
where $k \cdot S =0$ and $S^2 = - \vec{\mathds{P}}^2$.
One can also write $S^\mu$ explicitly in terms of the four-momenta 
of the $D_s$ meson and the $\tau$ lepton, using Eqs.~(\ref{eq:a7}) and (\ref{eq:a8}),     
\eq%
\begin{equation}
\label{eq:a9}
S^\mu = \frac{m_\tau}{k \cdot q} \, \bigl(p^\mu \, - \, k^\mu \, 
\frac{k \cdot p}{m_\tau^2} \, \bigr),   
\end{equation} 
\eq%
where $S^2 = -1$.
It follows that $\vec{\mathds P}^2 =1$, so that the $\tau$ lepton is $100\,\%$ polarised.

\vspace{.99cm}


\section{\label{app:Void} Optimal length for $D_s$ conversion to $\tau$ leptons}

We consider $\tau$ leptons produced in the decay of $D_s$ mesons and the consequent 
$\tau$ decay. Thus, there are two competing processes, production and decay of the $\tau$ lepton.

The probability of $\tau$ production in the process $D_s^+ \to \tau^+ \ \nu_{\tau}$ at the $D_s$ flight
length $x$ is
\eq
\begin{equation}
N_{\rm prod}=	{\cal B}_i
			\left(	1 - e ^ { - 	\,    x   \, / 
						\, T_D }
			\right),
\label{eq:Nprod}
\end{equation}
where
${\cal B}_i=\Gamma_i / \Gamma \approx 0.055$~\cite{PDG:2018} is the branching ratio of the current process,
$T_D=c\,\tau_D \gamma_D$ is the decay length of a boosted $D_s$ meson 
($c\,\tau_D\approx 150\,\mu{\rm m}$).
The probability of $\tau$ decay after a flight length of $L_{\rm v}-x$ is
\eq
\begin{equation}
N_{\rm dec}=	e^{-\,(L_{\rm v}-x)\,/\,T_\tau},
\label{eq:Ndec}
\end{equation}
where $T_\tau=c\,\tau_\tau\,\gamma_\tau$ is the decay length of a boosted $\tau$ lepton 
($c\,\tau_\tau \approx 87\,\mu{\rm m}$).

The conversion efficiency, that is the ratio of the number of $\tau$ leptons not decayed 
after the flight length $L_{\rm v}$, to the initial number of $D_s$ mesons of energy $ E_D$ is
\eq%
\begin{equation}
\eta(L_{\rm v}, E_D)  =	\dfrac{\int	\limits_{\e_\t^{min}}^{\e_\t^{max}}
					{\rm d}\e_\t
					\,\dfrac{\pd N_\t}{\pd\e_\t}
			    \ \,	\int	\limits_{0}^{L_{\rm v}}
					{\rm d}x
					\,\dfrac{\pd N_{\rm prod}}{\pd x}
					\ N_{\rm dec}\,(L_{\rm v}-x)
			}{	\int	\limits_{\e_\t^{min}}^{\e_\t^{max}}
					{\rm d}\e_\t
					\,\dfrac{\pd N_\t}{\pd\e_\t}
			},
\label{eq:Conversion}
\end{equation}
\eq%
where $\pd N_\t\,/\,\pd\e_\t$ is the $\tau$ spectrum. 
In a two-body decay, the $\tau$ spectrum is defined by the energy of the $D_s$ meson, but in our 
experimental setup, $\tau$ leptons are later captured into a channeling regime, and hence are collimated 
in angle and energy in a quite complicated way (see Fig.~\ref{fig:Sc_Coll_pxpy},\ref{fig:Ds-T_PE}).
Thus, in this case it is useful to find the extreme values of the conversion efficiency $\eta(L_{\rm v}, E_D)$ 
that correspond to the extreme values of the $\tau$ energy.

The energy range of $\tau$ leptons produced in a two-body decay of $D_s$ mesons in the relativistic case is
\eq%
\begin{equation}
\e_\tau \in	 E_D
		\left[\,\frac	{m_\t^2}
				{M_D^2}
			\,,\ 1\,
		\right].
\label{eq:E_T_range}
\end{equation}
\eq
 
Substituting Eqs.~\eqref{eq:Nprod} and \eqref{eq:Ndec} into Eq.~\eqref{eq:Conversion} and considering 
a mono-energetic $\tau$ spectrum of energy $\e_\t^*$, we obtain
\eq%
\begin{equation}
\eta^*(L_{\rm v}, E_D,\e_\t^*)  =	Br_i
				\ \frac	{e^{\,-L_{\rm v}\,/\,T_D} - e^{\,-L_{\rm v}\,/\,T_\t}}
						{T_D\,/\,T_\t-1}.
\label{eq:Conversion_Averaged}
\end{equation}
\eq

\begin{figure}[h]
\begin{center}
\includegraphics[width=0.64\textwidth]{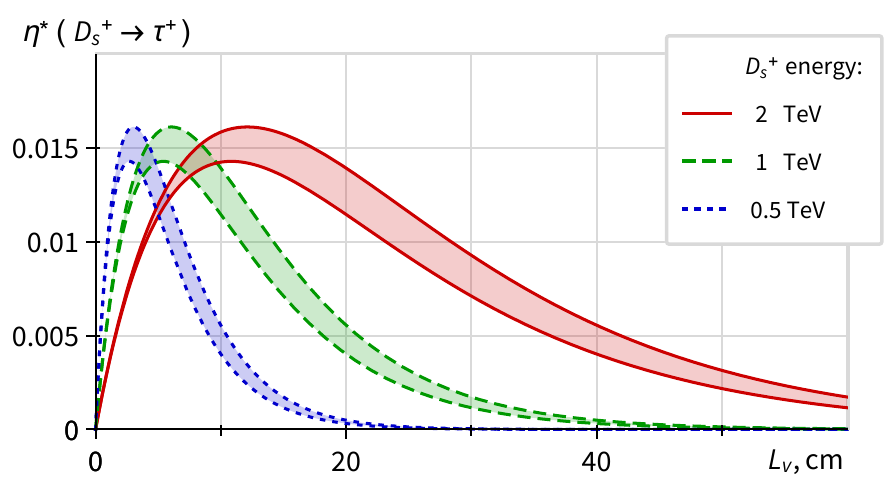}
\end{center}
\caption{$D_s^+ \to \tau^+$ conversion efficiency as a function of the flight length. }
\label{fig:Void}
\end{figure}

Figure~\ref{fig:Void} presents the conversion efficiency as a function of the path length for three  
$D_s$ energies, listed in the figure.
The width of the curves represents energy spread of the $\tau$ lepton. 
The upper and lower edges correspond to upper and lower limits of the $\tau$ energy 
given in Eq.~\eqref{eq:E_T_range}.

The optimal length for the $D_s$ conversion to $\tau$ leptons strongly depends on the $D_s$ spectrum.
For the energy range $1\,{\rm TeV}< E_D<2\,{\rm TeV}$, the optimal length is around $10\,{\rm cm}$.
Considering conversion lengths $L_{\rm v}>40\,{\rm cm}$ would essentially reduce the part of the spectra to
$E_D<2\,$TeV.

\newpage


\section*{Acknowledgments}

This research was partially conducted in the scope of the IDEATE International Associated Laboratory (LIA).
A.Yu.K. and A.S.F. acknowledge partial support by the National Academy of Sciences of Ukraine (project KPKVK 6541230, and project no. Ts-3/53-2018) and the Ministry of Education and Science of Ukraine (project no. 0117U004866).


\end{document}